\documentclass[pra,aps,twocolumn,superscriptaddress,showpacs,showkeys]{revtex4-2}

\usepackage[utf8]{inputenc}
\usepackage{graphicx}
\usepackage{dcolumn}
\usepackage{bm}
\usepackage{color} 
\usepackage{epsfig}
\usepackage{bbold}
\usepackage{url}
\usepackage{subfig}
\usepackage{lineno}
\usepackage{comment}
\usepackage{booktabs} 
\usepackage[colorlinks=true,citecolor={blue},urlcolor={red},linkcolor={blue}]{hyperref}
\usepackage{ulem} 

\begin{document}

\title{Quantum
Electrodynamics Coupled-Cluster Theory: Exploring Photon-Induced Electron Correlations}

\author{Himadri Pathak}
\email{himadri.pathak@pnnl.gov}
\affiliation{Advanced Computing, Mathematics, and Data Division, Pacific Northwest National Laboratory, Richland, Washington 99354, United States}

\author{Nicholas P. Bauman}
\affiliation{Physical Sciences Division, Pacific Northwest National Laboratory, Richland, Washington 99354, United States}

\author{Ajay Panyala}
\affiliation{Advanced Computing, Mathematics, and Data Division, Pacific Northwest National Laboratory, Richland, Washington 99354, United States}

\author{Karol Kowalski}
\email{karol.kowalski@pnnl.gov}
\affiliation{Physical Sciences Division, Pacific Northwest National Laboratory, Richland, Washington 99354, United States}

\date{\today}
\begin{abstract}
We present our successful implementation of the quantum electrodynamics coupled-cluster method with single and double excitations (QED-CCSD) for electronic and bosonic amplitudes, covering both individual and mixed excitation processes within the ExaChem program package, which relies on the Tensor Algebra for Many-body Methods (TAMM) infrastructure. TAMM is a parallel heterogeneous tensor library designed for utilizing modern computing platforms, from laptops to leadership-class computing resources.
This developed computational framework extends the traditional CCSD method to incorporate the intricate interplay between electronic and bosonic degrees of freedom, providing a comprehensive description of quantum phenomena. We discuss theoretical foundations, algorithmic details, and numerical benchmarks to demonstrate how the integration of bosonic degrees of freedom alters the electronic ground state.
The interactions between electrons and photons within an optical cavity are modeled using the Pauli-Fierz Hamiltonian within the dipole approximation in the length gauge.
The integration of QED effects within the CCSD framework contributes to a more accurate and versatile model for simulating complex quantum systems, thereby opening avenues for a better understanding, prediction, and manipulation of various physical phenomena.

\end{abstract}
\maketitle

\section{Introdcution}
The quest for precise and comprehensive quantum mechanical models to describe complex systems has catalyzed the evolution of sophisticated methodologies within computational chemistry and quantum physics~\cite{helgaker2013molecular, meyer2009multidimensional, grant2007relativistic, dyall2007introduction}. Advancements in high-performance computing now enable large-scale applications of these highly accurate techniques beyond model chemical systems~\cite{chang2023simulations, kowalski2021nwchem}.

Polariton chemistry has emerged as a captivating field due to its potential to alter chemical structures, properties, and reactivity through profound interactions among molecular electronic, vibrational, and rovibrational states~\cite{ruggenthaler2018quantum, hubener2017creating, byrnes2014exciton, basov2016polaritons, latini2019cavity, Thomas, thomas2, rubio_review}. While several research groups within the quantum optics community have devised model Hamiltonians to encapsulate the core aspects of polaritonic physics~\cite{feist1, feist2, Walther_2006, ribeiro2018polariton, campos2019resonant, gallego2024coherent}, a holistic quantitative theoretical understanding of molecular polaritons demands the equal treatment of both matter and photon interactions with quantum mechanical precision. In scenarios where molecular electronic interactions are strongly or ultrastrongly coupled to one or a few molecules, the application of \textit{ab initio} quantum chemistry techniques proves beneficial. This innovative approach, known as \textit{ab initio} cavity quantum electrodynamics, seamlessly integrates the molecular electronic degrees of freedom with the principles of cavity quantum electrodynamics governing photon behavior. The emergence and refinement of various \textit{ab initio} cavity quantum electrodynamics methodologies have gained significant traction in recent times.

Currently, the landscape of quantum electrodynamics extensions to various electronic structure methods includes density functional theory (QEDFT~\cite{rubio_chemrev, PhysRevA.84.042107, PhysRevA.90.012508, PhysRevLett.115.093001, doi:10.1021/acsphotonics.7b01279, doi:10.1080/00018732.2019.1695875, 10.1063/5.0021033} and QED-DFT~\cite{doi:10.1021/acs.jpca.2c07134, 10.1063/5.0095552, doi:10.1021/acs.jpca.3c01842}), real-time~\cite{PhysRevA.84.042107, PhysRevA.90.012508, PhysRevLett.110.233001, flick2017atoms, PhysRevB.98.235123, 10.1063/5.0123909} and linear-response formulations~\cite{doi:10.1021/acs.jpca.2c07134, 10.1063/5.0057542, 10.1063/5.0082386} of QED time-dependent density functional theory (QED-TDDFT), configuration interaction (QED-CIS)~\cite{10.1063/5.0091953}, cavity QED extensions of second-order Møller-Plesset perturbation theory, algebraic diagrammatic construction~\cite{10.1063/5.0142403, doi:10.1021/acs.jctc.3c01166}, variational QED-2-RDM methods~\cite{PhysRevA.106.053710}, and diffusion Monte Carlo~\cite{PhysRevA.109.032804}. Notably, the first formulation of QED-CASCI, both in the photon-number basis and the coherent-state basis, has recently been introduced~\cite{vu2024cavity}.

The time-independent coupled-cluster (CC) method, along with its myriad extensions~\cite{coester58_421, coester60_477, cizek66_4256, paldus72_50, kummel2003biography, RevModPhys.79.291, mukherjee1989use, krylov2008equation, piecuch2002recent, crawford2000introduction}, has proven invaluable in addressing electron correlation across a spectrum of problems spanning both ground and excited states. The ground-state CC method exhibits size-extensivity across all levels of excitation operator truncation and scales polynomially with the number of active orbitals. These attributes render CC methods particularly attractive compared to alternative electron correlation techniques, striking a harmonious balance between computational efficiency and accuracy. Furthermore, the results can be systematically improved by encompassing more correlated determinantal spaces. A hallmark feature of CC methods lies in their use of an exponential parameterization for the correlated ground-state wavefunction $\left|\Psi\right>$: $\left|\Psi\right>=e^{T}\left|\Phi\right>$, where $T$ denotes the cluster operator and $|\Phi\rangle$ is a reference wavefunction, typically—but not exclusively—a Hartree-Fock wavefunction.
\begin{table*}[!ht]
\centering
\caption{Comparison of QED-HF ground state energy and the correlation energy calculated at the \text{CCSD(2, 2)} level for the water molecule using various basis sets. Calculations were performed with the experimental geometry, with \(\lambda_z\) set to 0.1 and \(\Omega_{cav}\) set to 3.0 eV.}
\begin{ruledtabular}
\begin{tabular}{lcccc}
\textbf{Basis} &  \textbf{QED-HF} & & \textbf{CCSD(2, 2)}   \\
\cline{2-3}\cline{4-5} \\
               & this work & Ref.~\cite{flick_jacs}& this work & Ref.~\cite{flick_jacs}\\   
\midrule    
cc-pVDZ     & $-76.007\,2646$ & $-76.007\,2646$ &  $-0.216\,6871$& $-0.216\,6871$\\
aug-cc-pVDZ &$-76.020\,3868$ & $-76.020\,3868$  &$-0.233\,9280$ & $-0.233\,9280$\\
cc-pVTZ &$-76.036\,8309$ & $-76.036\,8309$  &$-0.285\,2382$& $-0.285\,2382$\\
aug-cc-pVTZ &$-76.039\,6515$ & $-76.039\,6515$  &$-0.292\,9503$& $\dots$\\
\end{tabular}
\end{ruledtabular}
\label{water_result}
\end{table*}
\begin{table}[!h]
\centering
\caption{Comparison of total energies (in Hartree) for the \( H_2 \) molecule across various basis sets using an experimental bond length of 0.74 \AA{} with \(\lambda_z = 0.05\) and \(\Omega_{cav} = 20.0 \) eV. The molecule is oriented in the \( z \)-direction with one of the hydrogens placed at the (0, 0, 0) coordinate so that there is a non-zero dipole moment in the direction of polarization.
}
\label{tab:fci_ccsd}
\begin{ruledtabular}
\begin{tabular}{rcc}
\textbf{Basis} &  \textbf{QED-CCSD(2,2)} & \textbf{QED-FCI}~\cite{vu2024cavity}  \\
\cline{2-3}\\
\hline
cc-pVDZ        & $-$1.162\,4643    &$-$1.162\,4643       \\
aug-cc-pVDZ    & $-$1.163\,6756        & $-$1.163\,6756  \\
cc-pVTZ        &   $-$1.171\,3895       & $-$1.171\,3895  \\
aug-cc-pVTZ    &    $-$1.171\,7020       &  $\dots$              \\   
\end{tabular}
\end{ruledtabular}
\end{table}
\begin{figure}[h]
\centering
 \includegraphics[width=0.5\textwidth]{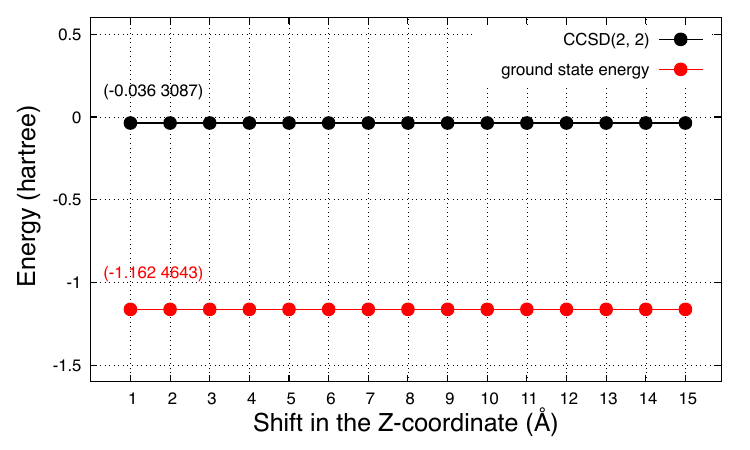}
\caption{
Correlation energy computed at the CCSD(2, 2) level and total ground-state energy as a function of the shift in the Z-coordinate, with the SCF energy being -1.1261\,556, which is also origin-invariant. The computations were performed with a shift in the Z-coordinate for the \( H_2 \) molecule using the aug-cc-pVDZ basis set. The calculations were carried out with \(\lambda_z = 0.05\) and \(\Omega_{cav} = 20.0 \) eV. The molecule is oriented along the \( z \)-axis, with one hydrogen positioned at the (0, 0, 0) coordinate, resulting in a non-zero dipole moment aligned with the polarization direction.
}
\label{fig:origin_invariance}
\end{figure}

\begin{table*}[!ht]
\centering
\caption{Comparison of QED-HF ground state energy and the correlation energy calculated at various \text{CCSD(m, n)} levels for the Malonaldehyde molecule using cc-pVDZ and aug-cc-pVDZ basis sets. The calculations were performed with three different polarizations, \(\lambda_x\), \(\lambda_y\), and \(\lambda_z\), set to 0.1, and \(\Omega_{\text{cav}}\) set to 3.0 eV.}
\label{tab:malonaldehyde}
\begin{ruledtabular}
\begin{tabular}{lrrrrrrrrr}
\textbf{Method} & \multicolumn{2}{c}{\textbf{X-polarization}} & \multicolumn{2}{c}{\textbf{Y-polarization}} & \multicolumn{2}{c}{\textbf{Z-polarization}}\\
\cline{2-3} \cline{4-5} \cline{6-7}\\
                & \textbf{cc-pVDZ} & \textbf{aug-cc-pVDZ} & \textbf{cc-pVDZ} & \textbf{aug-cc-pVDZ} & \textbf{cc-pVDZ} & \textbf{aug-cc-pVDZ}\\ 
\midrule    
QED-HF          & $-265.559\,2463$ & $-265.575\,7048$
                & $-265.557\,9089$ & $-265.574\,9429$
                & $-265.577\,3076$ & $-265.593\,1532$\\
QED-CCSD(2,\,0) & $-0.817\,9788$ & $-0.857\,4507$
                & $-0.820\,9197$ & $-0.859\,2885$
                & $-0.816\,3715$ & $-0.857\,8009$\\
QED-CCSD(2,\,1) & $-0.826\,4511$ & $-0.866\,8022$
                & $-0.828\,9601$ & $-0.867\,9947$
                & $-0.820\,1303$ & $-0.863\,5767$\\
QED-CCSD(2,\,2) & $-0.827\,2880$ & $-0.867\,8472$
                & $-0.829\,6532$ & $-0.868\,8461$
                & $-0.820\,2708$ & $-0.863\,9477$\\
\end{tabular}
\end{ruledtabular}
\end{table*}
\begin{table*}[t]
\centering
\caption{Comparison of QED-HF ground state energy and the correlation energy calculated at various \text{CCSD(m, n)} levels for the Aminopropenal molecule using cc-pVDZ and aug-cc-pVDZ basis sets. The calculations were performed with three different polarizations, \(\lambda_x\), \(\lambda_y\), and \(\lambda_z\), set to 0.1, and \(\Omega_{\text{cav}}\) set to 3.0 eV.}
\label{tab:aminopropenal}
\begin{ruledtabular}
\begin{tabular}{lrrrrrr}
\textbf{Method} & \multicolumn{2}{c}{\textbf{X-polarization}} & \multicolumn{2}{c}{\textbf{Y-polarization}} & \multicolumn{2}{c}{\textbf{Z-polarization}}\\
\cline{2-3} \cline{4-5} \cline{6-7}
                & \textbf{cc-pVDZ} & \textbf{aug-cc-pVDZ} & \textbf{cc-pVDZ} & \textbf{aug-cc-pVDZ} & \textbf{cc-pVDZ} & \textbf{aug-cc-pVDZ} \\ 
\midrule    
QED-HF          & $-245.724\,0393$ & $-245.740\,0942$
                & $-245.721\,9344$ & $-245.738\,4659$
                & $-245.743\,1042$ & $-245.758\,3281$ \\
QED-CCSD(2,\,0) & $-0.808\,3272$   & $-0.846\,4789$
                & $-0.812\,2811$   & $-0.849\,1371$
                & $-0.807\,0122$   & $-0.847\,1832$ \\
QED-CCSD(2,\,1) & $-0.817\,6584$   & $-0.856\,7412$
                & $-0.821\,0162$   & $-0.858\,6250$
                & $-0.811\,0344$   & $-0.853\,5370$ \\
QED-CCSD(2,\,2) & $-0.818\,6910$   & $-0.858\,0266$
                & $-0.821\,8357$   & $-0.859\,6684$
                & $-0.811\,1960$ & $-0.853\,9951$ \\
\end{tabular}
\end{ruledtabular}
\end{table*}

Two generalizations of coupled-cluster (CC) theory for incorporating quantum electrodynamics (QED) effects emerged around the same time. Mordovina {\it et al.}~\cite{mordovina} introduced the polaritonic coupled-cluster theory, which used an exponential parameterization of the ground-state polaritonic wavefunction. This parameterization included single and double electronic transition operators, photon creation operators, and combined electron transition and photon creation operators. They applied this method, along with QED full configuration interaction (CI), to study strong coupling between a single photon mode and a four-site Hubbard model. Notably, they used nilpotent operators instead of conventional boson creation operators, resulting in a linear parameterization of the photon space.
In contrast, the QED-CCSD-1 model by Haugland {\it et al.}~\cite{rubio_prx} adopted a similarly complex exponential parameterization but used conventional (non-nilpotent) boson creation operators. They applied this model to address strong coupling in an ab initio molecular Hamiltonian.

Several research teams have developed and applied similar QED-CC approaches to study the impact of cavity effects on various ground-state properties. DePrince~\cite{10.1063/5.0038748} used QED-CCSD-1 to show how strong coupling alters electron affinities in sodium halide compounds significantly. However, cavity effects had less influence on ionization potentials in these systems.

Pavo\v{s}evi\'{c} and Flick~\cite{doi:10.1021/acs.jpclett.1c02659} went further into cavity interactions' effects on electron affinities by employing a unitary formulation of QED-CCSD-1. They used the variational quantum eigensolver (VQE) algorithm~\cite{tilly2022variational} on a quantum computer for this purpose. Additionally, they expanded the framework to include up to two photon creation operators, alongside single and double electronic excitations, which they called QED-CCSD-2.

These foundational studies paved the way for further exploration of ionization characteristics in QED environments by Riso {\it et al.}~\cite{10.1063/5.0091119} underlining the critical importance of accurately treating the ionized electron.

Beyond the investigations focused on ionization and electron attachment, numerous studies have utilized QED-CC methodologies to examine the potential impact of vacuum fluctuations within chemical scenarios. It is crucial to emphasize that these studies concern alterations in the ground states of cavity-embedded systems, rather than inducing transitions or generating polariton states by introducing photons into the cavity.

Pavo\v{s}evi\'{c} {\it et al.}~\cite{flick_jacs} employed a non-unitary QED-CCSD-2 approach to demonstrate that strong coupling can lead to significant changes in proton transfer reaction barrier heights. The authors also introduced an approximation to QED-CCSD-2, where single electron transitions are paired with up to two photon creation operators, while double electron transitions are associated with single photon creation operators (referred to as QED-CCSD-21). This QED-CCSD-21 model bears resemblance to the method developed by White {\it et al.}~\cite{white}, which is designed to simulate electron-phonon interactions.

Pavo\v{s}evi\'{c} and Rubio~\cite{10.1063/5.0095552} incorporated QED-CCSD-1 into an embedding protocol. This approach focuses on a portion of a cavity-embedded molecular system using QED-CC, while the remaining part of the system is analyzed using either QED-DFT or QED-HF, termed as "QED-CC-in-QED-SCF". Assuming that electron-photon correlations are localized within the embedded region, this novel protocol could potentially alleviate the computational complexities inherent in the many-body ab initio cavity QED framework.
We refer to an excellent review article by DePrince \textit{et al.} \cite{deprince_review} for a more elaborate discussion on various \textit{Ab initio} methods for polariton chemistry.

In this article, we present our implementation of the QED-CC method using the Pauli-Fierz Hamiltonian. Our approach employs a coherent state basis and incorporates both single and double excitations, covering both individual and mixed excitations. This implementation is carried out on the ExaChem platform~\cite{doecode_108784}, a parallel heterogeneous computing platform. ExaChem is built upon the Tensor Algebra for Many-body Methods (TAMM)~\cite{tamm}, a specialized tensor library designed explicitly for the development of quantum chemistry applications and tailored for modern computing platforms, applicable to both CPUs and various GPUs.

The introduction provides an overview of the rationale behind integrating QED effects into the CCSD framework. It also highlights the significant advancements made in enhancing the accuracy of quantum simulations in prior research.
The remainder of this paper is as follows: In Sec.~\ref{method}, we briefly describe the QED-CC method within the single and double excitation approximation for both electronic and bosonic degrees of freedom. Sec.~\ref{exa} provides a brief overview of the details of the implementation within the ExaChem~\cite{doecode_108784} program package. Sec.~\ref{compdetail} and Sec.~\ref{results} are reserved for details of computational variables and parameters, and subsequent result and discussion section that includes parallel performance, before making final comments in Sec.~\ref{conclusion}.
We consistently use atomic units throughout this article unless otherwise specified.
\section{Method}\label{method}
\subsection{Coupled-Cluster theory for Electrons}
We consider a system of \( N \) electrons described by the electronic Hamiltonian \( H^e \) given by:
\begin{eqnarray}
H^e &=& \sum_{i=1}^{N} h(r_i, p_i) + \sum_{i=1}^{N-1} \sum_{j=i+1}^{N} \frac{1}{\left| r_i - r_j \right|},
\end{eqnarray}
where \( r_i \) and \( p_i \) represent the position and canonical momentum of electron \( i \), respectively. The second quantized form of the Hamiltonian is given by:

\begin{eqnarray}
H^e &=& h_{\nu}^{\mu} E_{\nu}^{\mu} + \frac{1}{2} u_{\nu \lambda}^{ \mu \gamma} E_{\nu \lambda}^{ \mu \gamma}\label{e-Hamiltonian2q},
\end{eqnarray}
where \( E_{\nu}^{\mu} = c^\dagger_{\mu} c_{\nu} \) and 
$E_{\nu \lambda}^{ \mu \gamma} = c_{\mu}^\dagger c_{\gamma}^\dagger c_{\lambda} c_{\nu}$
 with \( c^\dagger_{\mu} \) (\( c_{\mu} \)) being a creation (annihilation) operator in a complete, orthonormal set of \( 2n_{\text{basis}} \) spin-orbitals \(\{\psi_{\mu}\}\), where \( n_{\text{basis}} \) is the number of basis functions,

\begin{eqnarray}
h_{\nu}^{\mu} &=& \int dx_1 \psi^*_{\mu}(x_1) h(r_1, p_1) \psi_{\nu}(x_1),
\end{eqnarray}
\begin{eqnarray}
u^{\mu \gamma}_{\nu \lambda} &=& \int \int dx_1 dx_2 \frac{\psi^*_{\mu}(x_1) \psi^*_{\gamma}(x_2) \psi_{\nu}(x_1) \psi_{\lambda}(x_2)}{\left| r_1 - r_2 \right|},
\end{eqnarray}
where \( x_i = (r_i, \sigma_i) \) is a composite spatial-spin coordinate.

The ground-state wave function in the CC method~\cite{coester58_421, coester60_477, cizek66_4256, paldus72_50, kummel2003biography} is defined as
\begin{eqnarray}
|\Psi\rangle &=& e^T |0^e\rangle,
\end{eqnarray}
where \( |0^e\rangle \) denotes a single Slater determinant, which is typically, but not always, the \( N \)-electron Hartree-Fock reference determinant. \( T \) represents the cluster operator. When applied to the reference determinant, it generates excited determinants that encompass correlated determinantal configurations, thereby accounting for electron correlations.

The fermionic cluster operators are expressed as follows:
\begin{eqnarray}
T &=& T_1 + T_2 + \dots, \nonumber \\
  &=& \sum_{i,a} t_{i}^{a} E_{i}^{a} + \sum_{a<b, i<j} t_{ij}^{ab} E_{ij}^{ab} + \dots. \label{eq:cluster_operator}
\end{eqnarray}
Here, the excitation operators \( T_1 \), \( T_2 \), and so on, generate singly and doubly excited determinants, and so forth, when acting on the reference function $|0^e\rangle$.

The indices \( i, j \) and \( a, b \) refer to the occupied and virtual spin-orbitals, respectively, with respect to the reference determinant. Hereafter, we will simply refer to them as orbitals.

Premultiplying the electronic Schr\"odinger equation involving the CC wavefunction \( H^e e^T |0^e\rangle = E e^T |0^e\rangle \) by the nonsingular operator \( e^{-T} \) and projecting it onto the basis of excited determinants \( |\Phi_{ij\dots}^{ab\dots} \rangle \) relative to the \( N \)-electron Hartree-Fock determinant, we obtain the equations for the amplitudes:
\begin{eqnarray}
\langle \Phi_{ij\dots}^{ab\dots} | e^{-T} H^e e^{T} |0^e \rangle = 0.
\end{eqnarray}
These are coupled, nonlinear, simultaneous equations that are solved iteratively. Finally, the ground-state energy can be obtained from the energy expression of Eq.~\ref{cc_energy} using the solved amplitudes obtained within a certain given threshold.
\begin{eqnarray}
E_{\text{CC}} &=& \langle 0^e | e^{-T} H^e e^{T} |0^e\rangle\label{cc_energy}.
\end{eqnarray}

\subsection{Unified Coupled-Cluster Framework for Electron-Photon Interactions}
In general, the interactions between molecules and photons within an optical cavity are typically modeled using the Pauli-Fierz Hamiltonian~\cite{ruggenthaler2018quantum, cohen1997photons, spohn2004dynamics, rokaj2018light}. To incorporate the optical cavity, we couple the electronic system to a single photon mode. We utilize the dipole approximation, which is justified by the significant disparity between the wavelength of the photon mode and the dimensions of our molecular system. Additionally, we adopt the coherent state basis to facilitate our implementation. Given these assumptions, and following the prescription of Haugland {\it et al.}~\cite{rubio_prx}, the Pauli-Fierz Hamiltonian takes the following form:
\begin{eqnarray}
H &=& H_e + \Omega_{\text{cav}} b^\dag b - \sqrt{\frac{\Omega_{\text{cav}}}{2}}
(\lambda\cdot ({\mathbf{d}}-\langle {\mathbf{d}}\rangle))(b^\dag +b)\nonumber\\
&&+\frac{1}{2}(\lambda\cdot ({\mathbf{d}}-\langle {\mathbf{d}}\rangle))^2
\end{eqnarray}

Here, \(H_e\) is the standard molecular electronic Hamiltonian within the Born-Oppenheimer approximation defined in Eq.~\ref{e-Hamiltonian2q}. The second term in the Hamiltonian represents the photonic part, modeled by a harmonic oscillator with \( \Omega_{\text{cav}} \) as the fundamental frequency, and \( b^\dagger \) and \( b \) are bosonic creation and annihilation operators, respectively.

The third term is the bi-linear coupling that couples fermionic and bosonic degrees of freedom with a coupling strength vector \( \lambda \).
The value of \(\lambda\) is related to the field strength of the photon mode. The coupling strength \(\lambda\) can be defined by an effective volume \(V_{\text{eff}}\). Specifically, \(\lambda\) is given by \(\lambda = \sqrt{\frac{1}{{\epsilon_0}V_{\text{eff}}}}\)~\cite{climent2019plasmonic, feist1}, where \(\epsilon_0\) is the vacuum permittivity.

The last term in the Hamiltonian is the dipole self-energy, which ensures that the Hamiltonian is bounded from the below and is origin-independent~\cite{doi:10.1021/acsphotonics.9b01649, PhysRevA.97.043820, rokaj2018light, PhysRevLett.125.123602}.
\(\mathbf{d}\) is the molecular dipole operator given by \( \mathbf{d} = d_q^pE_q^p \) with \( d_q^p = \langle p|{\mathbf{d}}_e+\frac{{\mathbf{d}}_{\text{nuc}}}{N_e}|q\rangle \).
where ${\bm{d_e}}$ is electronic dipole-moment and   
\begin{eqnarray}
{\mathbf{d}}_{\text{nuc}} &=& \sum_A Z_A \mathbf{R}_A
\end{eqnarray}
is the nuclear component. 
Here, \( Z_A \) is the nuclear charge, \( \mathbf{R}_A \) is the nuclear displacement vector of atomic nucleus \( A \), and \( N_e \) is the total number of electrons under consideration.
The nuclear dipole moment does not contribute to the Hamiltonian in the coherent-state basis. However, it appears in the wavefunction through the coherent-state transformation. For details, see Ref.~\cite{rubio_prx}.

This Hamiltonian is origin invariant for neutral systems; however, it is not origin invariant for charged systems, where the origin dependence arises from the dipole operator~\cite{rubio_prx}. The error or change in the ground state energy due to the origin dependence can be minimized by placing the molecule at the center of its mass.

The starting point for the QED-CC method is a QED-HF reference wavefunction, which extends the electronic HF method. 
The QED reference wavefunction is defined as~\cite{rubio_prx, rivera2019variational},
\begin{equation}
|R\rangle = |0^e\rangle \otimes |0^{ph}\rangle    
\end{equation}
This is a direct product of the electronic ($|0^e\rangle$) and photonic ($|0^{ph}\rangle$) reference wavefunctions.  

The wavefunction in the QED-CC method is defined as~\cite{rubio_prx},
\begin{equation}
|\Psi_{\text{QED-CC}}\rangle = e^{T(e, ph)}|R\rangle     
\end{equation}
The notation $e$ and $ph$ within the parentheses of the $T$ operator correspond to electronic and photonic excitation manifolds from the reference wavefunction.
The cluster operator $T= \sum_{\mu, n}t_{\mu, n}a^{\mu}(b^\dag)^n$ consists of unknown parameters that are determined by solving a set of nonlinear equations~\cite{bartlett2007coupled, rubio_prx},
\begin{equation}
\langle R|a_{\mu}(b)^ne^{-T}He^{T}|R\rangle = \sigma_{\mu, n} = 0   
\end{equation}
Here, \(a^{\mu}=a_{\mu}^{\dag}=\{E_i^a, E_{ij}^{ab}, \dots\}\) represents the electronic excitation operator. The index $\mu$ indicates the electronic excitation rank, and $n$ denotes the number of photons.

\section{Implementation within ExaChem}\label{exa}
The way the dipole self-energy term is applied varies across different studies due to different treatments of the squared electric dipole operator. To understand these discrepancies, let's begin by expanding the dipole self-energy operator as follows:
\begin{eqnarray}
\frac{1}{2} [\lambda\cdot (\bm{d} - \langle \bm{d} \rangle)]^2=
\frac{1}{2} (\lambda\cdot {\bm{d}})^2-(\lambda\cdot {\bm{d}})(\lambda\cdot {\langle \bm{d}\rangle})\nonumber\\
+ \frac{1}{2}(\lambda\cdot\langle {\bm d}\rangle)^2\label{sqd}
\end{eqnarray}
The square of the electric dipole operator, which is the first term on the right-hand side of Eq. \ref{sqd}, can be expanded into contributions from one-electron and two-electron terms as follows:
\begin{eqnarray}
(\lambda\cdot {\bm{d}})^2 &=& \sum_{i\neq j}   [\lambda\cdot {\bm d}(i)]
[\lambda\cdot {\bm d}(j)]+\sum_{i} [\lambda\cdot {\bm d}(i)]^2\\
&=&\lambda^2\sum_{\mu\nu\lambda\sigma}{\bm d}_{\nu}^\mu {\bm d}_{\sigma}^\lambda E_{\sigma\nu}^{\mu \lambda}-\lambda^2 \sum_{\mu\nu}
Q_\nu^\mu E_\nu^\mu
\end{eqnarray}
The symbols ${\bm d}_\nu^\mu$ and ${Q}_\nu^\mu$ represent electric dipole and electric quadrupole integrals, which have the form:
\begin{eqnarray}
\langle \mu|d_e|\nu \rangle &=& -\int \phi_\mu^*(\mathbf{r_1}) \mathbf{r_1} \phi_\nu(\mathbf{r_1}) \, d\mathbf{r_1} 
\end{eqnarray}
\begin{eqnarray}
\langle \mu|Q|\nu \rangle= -\int \phi_\mu^*(\mathbf{r_1}) \mathbf{r_1 r_2} \phi_\nu(\mathbf{r_2}) \, d\mathbf{r_1}  d\mathbf{r_2}  
\end{eqnarray}
On the other hand, numerous studies interpret the second-quantized form of the square of the electric dipole operator as the product of second-quantized electric dipole operators~\cite{rubio_prx}, which leads to
\begin{eqnarray}
(\lambda\cdot {\bm{d}})^2
&=&\lambda^2\sum_{\mu\nu\lambda\sigma}{\bm d}_{\nu}^\mu {\bm d}_{\sigma}^\lambda E_{\sigma\nu}^{\mu \lambda}+\lambda^2 \sum_{\mu\nu}
\sum_\sigma {\bm d}_\sigma^\mu {\bm d}_\nu^\sigma E_\nu^\mu
\end{eqnarray}
Taking into account of the expansion of the square of the dipole operator,
we rewrite the Pauli-Fierz Hamiltonian as,
\begin{eqnarray}
H &=& H_e^\prime + \Omega_{\text{cav}} b^\dag b - \sqrt{\frac{\Omega_{\text{cav}}}{2}}
(\lambda\cdot ({\mathbf{d}}-\langle {\mathbf{d}}\rangle))(b^\dag +b)\nonumber\\
&&+\frac{1}{2}(\lambda\cdot\langle {\mathbf{d}}\rangle))^2\label{pfham}
\end{eqnarray}
Here, $H_e^\prime$ is 
\begin{eqnarray}
H_e^\prime= (h_q^p-\frac{1}{2} \lambda^2 \cdot Q_q^p-\langle{\bm d}\rangle \cdot {\bm d})E_q^p\nonumber\\
+\frac{1}{2}\lambda^2 \cdot (u_{qs}^{pr}+{\bm d}_{q}^{p}\otimes{\bm d}_{s}^{r})E_{qs}^{pr}  
\end{eqnarray}

When averaging over the photon vacuum state \( |0^{ph}\rangle \), the PF Hamiltonian of Eq.~\ref{pfham} takes the form for the QED-HF procedure as follows~\cite{rubio_prx}:
\begin{eqnarray}
\langle H\rangle_{0^{ph}} = 
H_e^\prime + \frac{1}{2}(\lambda\cdot\langle {\mathbf{d}}\rangle)^2
\end{eqnarray}
This procedure allows one to avoid optimizing photonic orbitals at the QED-HF level. With minor modifications to the existing electronic HF code, one can solve for the QED-HF reference wavefunction. The purely photonic part of the Hamiltonian and the bi-linear coupling term should be taken into account at the post-Hartree-Fock level.

The truncation of the cluster operator at a certain excitation rank $m$ and number of photons $n$ establishes the QED-CC hierarchy. 
Throughout this paper, we adopt the QED-CCSD(m, n) notation, which represents the highest degree of interactions involving $m$ electrons with $n$ photons. Below, these notations are presented in equation form:


\begin{widetext}
\begin{eqnarray}
T(e, ph) &=& \overbrace{\overbrace{\overbrace{T(1, 0) + T(2, 0)}^{\text{CCSD(2, 0)}} + T(0, 1) + T(1, 1) + T(2, 1)}^{\text{CCSD(2, 1)}} + T(0, 2) + T(1, 2) + T(2, 2)}^{\text{CCSD(2, 2)}} \\
&=& t_i^{a,0} E_i^a + t_{ij}^{ab, 0} E_{ij}^{ab} + t^{0, 1} b^\dag + t_i^{a, 1} E_i^a b^\dag + t_{ij}^{ab, 1} E_{ij}^{ab} b^\dag + t^{0, 2} b^\dag b^\dag + t_i^{a,2} E_i^a b^\dag b^\dag + t_{ij}^{ab,2} E_{ij}^{ab} b^\dag b^\dag 
\end{eqnarray}
\end{widetext}

We have devised three approximate schemes to understand how photon-electron interactions influence the electronic ground state. We denote these approximate schemes as CCSD(2, 0), CCSD(2, 1), and CCSD(2, 2). The CCSD(2, 0) involves performing electronic CCSD calculations with the QED Fock matrix elements and two-electron integrals that are transformed using the coefficient matrix elements obtained from the QED-HF procedure.

We can also obtain the same results as CCSD(2, 0) by setting \(\Omega_{\text{cav}} = 0\) in either CCSD(2, 1) or CCSD(2, 2). The parameter \(\Omega_{\text{cav}}\) is inversely proportional to the length of the cavity, \(L\) (\(\Omega_{\text{cav}} \propto \frac{1}{L}\)), meaning it is essentially performing QED-CCSD(m, n) simulations with a cavity of infinite length.

CCSD(2, 0) does not require any new amplitude tensor definitions; the electronic \(T_1\) and \(T_2\) are sufficient. However, the CCSD(2, 1) scheme requires a total of five kinds of amplitude tensors, namely \(T(1, 0)\), \(T(2, 0)\), \(T(0, 1)\), \(T(1, 1)\), and \(T(2, 1)\). The CCSD(2, 2) requires an additional three new tensors \(T(0, 2)\), \(T(1, 2)\), and \(T(2, 2)\). Out of these, \(T(0, 1)\) and \(T(0, 2)\) are rank-zero tensors or scalars. Since the bosonic part is a scalar, the mixed tensors \(T(1, 1)\), \(T(2, 1)\), \(T(1, 2)\), and \(T(2, 2)\) do not increase any dimensionality in the tensor definition. Therefore, the computational complexity for all the schemes remains the same as the electronic CCSD method of \(N^6\); however, these schemes increase the prefactors in the computational complexity.
As pointed out earlier, the nuclear dipole moment does not contribute to the Hamiltonian in the coherent-state basis. However, we observed that the inclusion of the \(\frac{{\mathbf{d}}_{\text{nuc}}}{N_e}\) term in the Hamiltonian results in more stable convergence at both the SCF and CC levels.

We have implemented the QED-CC method within the ExaChem program package~\cite{doecode_108784}, which is a freely distributed open-source software. ExaChem is built upon the Tensor Algebra for Many-body Methods (TAMM) infrastructure~\cite{tamm}.
TAMM is a massively parallel, heterogeneous tensor algebra library designed for scalable performance. It supports portable implementations of many-body methods on current and future exascale supercomputing platforms. TAMM enables users to specify and manage tensor distribution, memory, and operation scheduling, and supports both complex and mixed real-complex algebra. It utilizes Global Arrays~\cite{ga_paper} and MPI for scalable parallelization on distributed memory systems, along with optimized libraries for efficient intra-node execution on CPUs and accelerators.
For GPU execution, TAMM uses localized summation loops to minimize data transfer between GPUs and CPUs, thereby reduce CPU-GPU data transmission.

\section{Computational Details}\label{compdetail}
In this article, we report results for several closed-shell molecules, namely hydrogen, water, malonaldehyde, and aminopropenal, using various basis sets. For hydrogen and water molecules, we have used experimental geometries. The geometries of malonaldehyde and aminopropenal are taken from Ref.~\cite{flick_jacs} and were optimized at the electronic CCSD level using the cc-pVDZ~\cite{dunning1989gaussian} basis set.

The hydrogen molecule serves as a reference system for comparing the accuracy of our CCSD(2, 2) implementation with the QED-FCI method, as detailed in \cite{vu2024cavity}. We utilized cc-pVDZ~\cite{dunning1989gaussian}, aug-cc-pVDZ~\cite{kendall1992electron}, cc-pVTZ~\cite{dunning1989gaussian}, and aug-cc-pVTZ~\cite{kendall1992electron} basis sets for both hydrogen in the hydrogen molecule and hydrogen and oxygen in water. To validate our implementation further, we compared our results for water molecules with those obtained using the code by Flick et al. \cite{flick_jacs}.

For hydrogen, carbon, and oxygen in the malonaldehyde molecule, we employed the cc-pVDZ~\cite{dunning1989gaussian} and aug-cc-pVDZ~\cite{kendall1992electron} basis sets. Similarly, for the aminopropenal molecule, we used the cc-pVDZ~\cite{dunning1989gaussian} and aug-cc-pVDZ~\cite{kendall1992electron} basis sets for hydrogen, carbon, nitrogen, and oxygen atoms. We also report results for variations in ground-state energy by varying either \(\lambda\) or \(\Omega_{\text{cav}}\) while keeping the other constant, using aminopropenal as an example.

The necessary matrix elements for implementing the QED-CCSD(m, n) methods within the ExaChem~\cite{doecode_108784} infrastructure are taken from the libint2.9.0 integral engine. We used an integral threshold of \(10^{-20}\) and a linear dependence threshold of \(10^{-5}\) in all our simulations. At the SCF level, we employed a threshold of \(10^{-9}\) for density and \(10^{-8}\) Frobenius norm of the residual vector unless specifically reported. All calculations used a direct inversion in the iterative subspace (DIIS) of 5. None of the electrons were frozen in any of our calculations.

\section{Results and discussion}\label{results}
To verify the correctness of our implementation, we conducted various tests. One such test compared the results from our ExaChem implementation of the QED-CCSD(2, 2) method with the developmental version of the QED-CCSD-mn methods by Flick \cite{flick_jacs}, available as supplemental material in Ref.~\cite{flick_jacs}.
We used water molecule with experimental geometry as the test case across various basis sets, with \(\lambda_z=0.1\) and \(\Omega_{\text{cav}}=3.0\,\text{eV}\), ranging from cc-pVDZ to aug-cc-pVTZ. The choice of cavity parameters is arbitrary. The results are summarized in Table~\ref{water_result}.
The QED-CCSD-mn simulations were performed using a MacBook Pro with 16 GB of RAM having Apple M1 chip, which is insufficient to simulate even a water molecule using the aug-cc-pVTZ basis set, and therefore, it is not reported in the table.
We conducted many other tests by varying \(\Omega_{\text{cav}}\) or the direction of the polarization with various coupling strengths. In all cases, we achieved nine to ten digits of agreement after the decimal for the SCF method and seven to eight digits of agreement for the CCSD(2, 2), irrespective of the simulation conditions. The discrepancy beyond this limit is due to the fact that the two codes use different convergence algorithms. The code of Ref.~\cite{flick_jacs} uses a convergence threshold over difference in correlation energy between two successive iterations, whereas we use a threshold over the maximum of the Frobenius norm of the residual vectors.

Furthermore, we have compared our QED-CCSD(2, 2) results for a two-electron system with values computed using the coherent-state quantum electrodynamics version of full configuration interaction (CS-QED-FCI)\cite{vu2024cavity}. Theoretically, the coherent-state basis represents an infinite photon occupation.
However, CS-QED-FCI has the option to use finite photon occupation. The CS-QED-FCI results are reported with converged photon occupation.
We used hydrogen molecule at its experimental bond length of 0.74 \AA{}. The molecule is oriented in the \( z \)-direction with one of the hydrogens placed at the (0, 0, 0) coordinate so that there is a non-zero dipole moment in the direction of polarization. We have used coupling strength \(\lambda_z = 0.05\) and \(\Omega_{\text{cav}} = 20.0\,\text{eV}\) in all our comparisons. Here also, we employed cc-pVDZ to aug-cc-pVTZ basis sets for computing ground state energy using the QED-CCSD(2, 2) method. We obtained perfect agreement with QED-FCI~\cite{vu2024cavity} values wherever it was possible to compute. The results are reported in Tab~\ref{tab:fci_ccsd}

In Fig.~\ref{fig:origin_invariance}, we report the results of shifting the Z-coordinate and the ground-state energy computed at the QED-HF and QED-CCSD(2, 2) levels. The hydrogen molecule is oriented in the \(z\)-direction, with one of the hydrogen atoms placed at the center of the coordinate system to ensure a non-zero dipole moment in that direction. We employed \(\lambda_z = 0.05\) and \(\Omega_{\text{cav}} = 20.0\,\text{eV}\) using the aug-cc-pVDZ basis set. We gradually shifted the Z-coordinate starting with 1 \AA{} to 16 \AA{}. We observed no change in the QED-HF or QED-CCSD(2, 2) correlation energies. Therefore, we can conclude that our implementation is origin-invariant.

In Table~\ref{tab:malonaldehyde}, we present results for the malonaldehyde molecule using cc-pVDZ and aug-cc-pVDZ basis sets with QED-CCSD(m, n) schemes, considering either X-, Y-, or Z-polarization with a coupling strength of 0.1. In all cases, we used \(\Omega_{\text{cav}} = 3.0\) eV. The electronic SCF energy of the molecule is \(-265.6559200\) in the cc-pVDZ basis set and \(-265.6734631\) in the aug-cc-pVDZ basis set. The purely electronic correlation energy changes from \(-0.8024509\) to \(-0.8399609\) when switching from the cc-pVDZ basis set to the aug-cc-pVDZ basis set.

The QED-HF ground state energy for X-polarization is $-265.5592463$ with the cc-pVDZ basis set and $-265.5757048$ with the aug-cc-pVDZ basis set. For Y-polarization, the QED-HF ground state energy is $-265.5579089$ with the cc-pVDZ basis set and $-265.5749429$ with the aug-cc-pVDZ basis set. For Z-polarization, the QED-HF ground state energy is $-265.5773076$ with the cc-pVDZ basis set and $-265.5931532$ with the aug-cc-pVDZ basis set.

The QED-CCSD correlation energies for the (2,0) level are $-0.8179788$ (cc-pVDZ) and $-0.8574507$ (aug-cc-pVDZ) for X-polarization, $-0.8209197$ (cc-pVDZ) and $-0.8592885$ (aug-cc-pVDZ) for Y-polarization, and $-0.8163715$ (cc-pVDZ) and $-0.8578009$ (aug-cc-pVDZ) for Z-polarization. For the (2,1) level, the QED-CCSD correlation energies are $-0.8264511$ (cc-pVDZ) and $-0.8668022$ (aug-cc-pVDZ) for X-polarization, $-0.8289601$ (cc-pVDZ) and $-0.8679947$ (aug-cc-pVDZ) for Y-polarization, and $-0.8201303$ (cc-pVDZ) and $-0.8635767$ (aug-cc-pVDZ) for Z-polarization. For the (2,2) level, the QED-CCSD correlation energies are $-0.8272880$ (cc-pVDZ) and $-0.8678472$ (aug-cc-pVDZ) for X-polarization, $-0.8296532$ (cc-pVDZ) and $-0.8688461$ (aug-cc-pVDZ) for Y-polarization, and $-0.8202708$ (cc-pVDZ) and $-0.8639477$ (aug-cc-pVDZ) for Z-polarization.

The ground state energy obtained from QED-HF calculations is lower (more negative) when using the aug-cc-pVDZ basis set compared to the cc-pVDZ basis set across all polarizations. This suggests that the aug-cc-pVDZ basis set, being larger and more complete, provides a more accurate representation of the electronic structure. The correlation energy, which accounts for electron correlation effects beyond the mean-field approximation, also shows more negative values with the aug-cc-pVDZ basis set compared to the cc-pVDZ basis set. This trend is consistent across different levels of QED-CCSD (2,0), (2,1), and (2,2) and different polarizations. The correlation energy values indicate that the electron correlation is better captured with the more comprehensive aug-cc-pVDZ basis set. The results show slight variations in energies depending on the polarization direction (X, Y, or Z). For instance, the Z-polarization often results in slightly more negative QED-HF energies compared to X and Y polarizations, which could be indicative of specific interactions between the electronic structure of malonaldehyde and the electromagnetic field in that direction. Overall, the table highlights the importance of basis set selection and polarization in obtaining accurate quantum chemical calculations, with the aug-cc-pVDZ basis set generally providing better results.

In Table~\ref{tab:aminopropenal}, we present the results for the aminopropenal molecule using cc-pVDZ and aug-cc-pVDZ basis sets with QED-CCSD(m, n) schemes, considering X-, Y-, or Z-polarization with a coupling strength of 0.1. In all cases, \(\Omega_{\text{cav}} = 3.0\) eV. This molecule has a convergence difficulty when the polarization is applied along the Y-direction. All Y-polarization results are calculated using a threshold of \(10^{-6}\).

The QED-HF ground state energy for X-polarization is $-245.7240393$ with the cc-pVDZ basis set and $-245.7400942$ with the aug-cc-pVDZ basis set. For Y-polarization, the QED-HF ground state energy is $-245.7219344$ with the cc-pVDZ basis set and $-245.7384659$ with the aug-cc-pVDZ basis set. For Z-polarization, the QED-HF ground state energy is $-245.7431042$ with the cc-pVDZ basis set and $-245.7583281$ with the aug-cc-pVDZ basis set. The electronic SCF energy of the aminopropenal molecule is \(-245.8268167\) with the cc-pVDZ basis set and \(-245.8443871\) with the aug-cc-pVDZ basis set. The purely electronic correlation energy changes from \(-0.7917050\) to \(-0.82750579\) when switching from the cc-pVDZ basis set to the aug-cc-pVDZ basis set.

The QED-CCSD correlation energies for the (2,0) level are $-0.8083272$ (cc-pVDZ) and $-0.8464789$ (aug-cc-pVDZ) for X-polarization, $-0.8122811$ (cc-pVDZ) and $-0.8491371$ (aug-cc-pVDZ) for Y-polarization, and $-0.8070122$ (cc-pVDZ) and $-0.8471832$ (aug-cc-pVDZ) for Z-polarization. For the (2,1) level, the QED-CCSD correlation energies are $-0.8176584$ (cc-pVDZ) and $-0.8567412$ (aug-cc-pVDZ) for X-polarization, $-0.8210162$ (cc-pVDZ) and $-0.8586250$ (aug-cc-pVDZ) for Y-polarization, and $-0.8110344$ (cc-pVDZ) and $-0.8535370$ (aug-cc-pVDZ) for Z-polarization. For the (2,2) level, the QED-CCSD correlation energies are $-0.8186910$ (cc-pVDZ) and $-0.8580266$ (aug-cc-pVDZ) for X-polarization, $-0.8218357$ (cc-pVDZ) and $-0.8596684$ (aug-cc-pVDZ) for Y-polarization, and $-0.8111960$ (cc-pVDZ) and $-0.8539951$ (aug-cc-pVDZ) for Z-polarization.

The QED-HF calculations consistently yield lower (more negative) ground state energies when employing the aug-cc-pVDZ basis set compared to the cc-pVDZ basis set across all polarizations. This suggests that the aug-cc-pVDZ basis set, being more extensive and comprehensive, provides a more accurate representation of the electronic structure. Furthermore, the correlation energy, which accounts for electron correlation effects beyond the mean-field approximation, also demonstrates more negative values with the aug-cc-pVDZ basis set compared to the cc-pVDZ basis set. This trend persists across various levels of QED-CCSD (2,0), (2,1), and (2,2), as well as different polarizations. The correlation energy values indicate that the electron correlation is better captured with the more sophisticated aug-cc-pVDZ basis set.

The results exhibit slight energy variations depending on the polarization direction (X, Y, or Z). For instance, Z-polarization often yields slightly more negative QED-HF energies compared to X and Y polarizations, implying potential specific interactions between the electronic structure of aminopropenal and the electromagnetic field in that direction. Overall, the table underscores the importance of both basis set selection and polarization in achieving accurate quantum chemical calculations, with the aug-cc-pVDZ basis set generally providing superior results.
\begin{figure}[h]
\centering
 \includegraphics[width=0.5\textwidth]{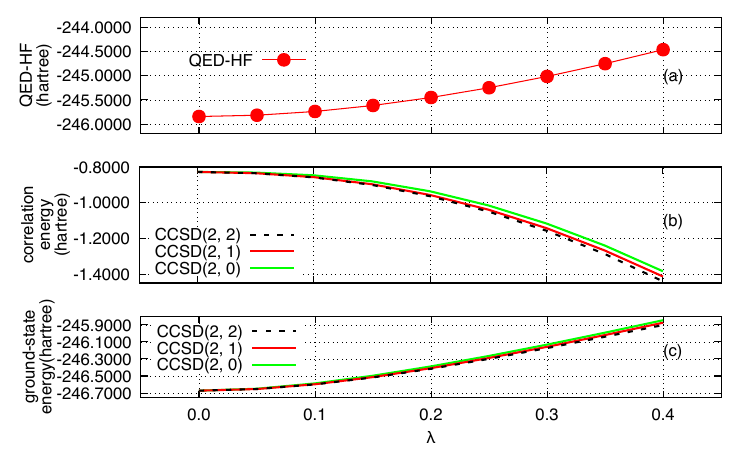}
\caption{Variation of QED-HF ground state energy (a), correlation energies using various CCSD(m, n) schemes (b), and total ground-state energy at the QED-CCSD(m, n) levels (c) with the variation in $\lambda_x$, while keeping $\Omega_{\text{cav}} = 3.0$ eV using the aug-cc-pVDZ basis set.}
\label{fig:lambda_variation}
\end{figure}

The results depicted in Fig.~\ref{fig:lambda_variation} show the variation of QED-HF ground state energy, correlation energies using different CCSD(m, n) schemes, and total ground state energy at the QED-CC levels with respect to the polarization parameter $\lambda_x$. These calculations were conducted using the aug-cc-pVDZ~\cite{kendall1992electron} basis set with a fixed cavity frequency of $\Omega_{\text{cav}}=3.0$ eV.

As $\lambda_x$ increases from 0 to 0.4, there is a consistent decrease in the QED-HF ground state energy. This indicates a stronger interaction between the molecular system and the electromagnetic field along the x-direction. When $\lambda_x = 0$, the system's behavior essentially reflects the effects outside of the cavity, or in other words, these effects are of purely electronic origin.

In contrast to the QED-HF energy, the correlation energy computed using various CCSD(m, n) schemes shows an opposite trend. There is an increase in the correlation energy with the increase in the electron-photon coupling strength. Among the three approximation schemes, the CCSD(2, 2) scheme exhibits the largest increase in correlation energy, while the CCSD(2, 1) scheme shows a moderate increase. This suggests that higher-order electron correlation effects become more pronounced with stronger coupling to the photon field.
Despite the opposing trends in QED-HF and QED-CCSD(m, n) correlation energies, the overall ground state energy at the QED-CCSD(m, n) level decreases with increasing $\lambda_x$. This implies that the total electronic ground state is significantly influenced by the degree of polarization. 
The observed decrease in ground state energy, despite the opposing trends in individual energy components, highlights the complex interplay between electronic and photonic interactions in the system. It underscores the necessity of using advanced quantum electrodynamic methods to accurately model these interactions. The results demonstrate the sensitivity of the molecular system's properties to external electromagnetic fields and polarization parameters, emphasizing the importance of precise modeling for reliable predictions of molecular properties.
In summary, the data reveal that while the QED-HF ground state energy decreases with increasing $\lambda_x$, the correlation energies increase, particularly for higher-order CCSD schemes. This results in an overall decrease in the total ground state energy at the QED-CCSD level, illustrating the nuanced effects of electron-photon coupling on the electronic structure. Such insights are crucial for developing accurate quantum chemical methods that account for the influence of external fields and polarization effects.
\begin{figure}[h]
\centering
 \includegraphics[width=0.5\textwidth]{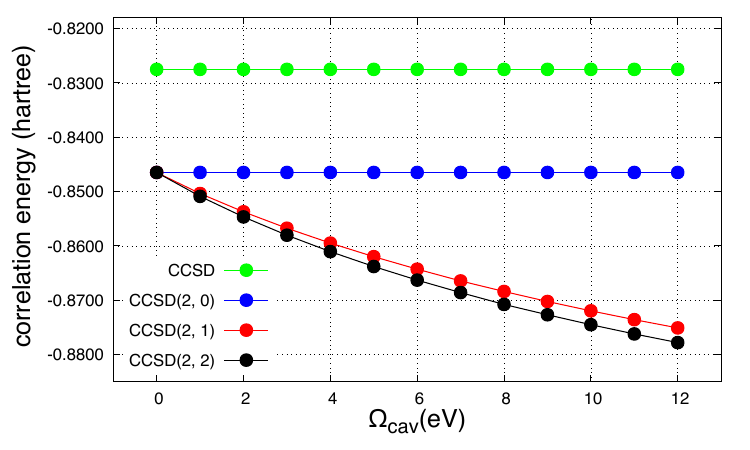}
\caption{
Variation of correlation energies using various CCSD(m, n) schemes as a function of $\Omega_{\text{cav}}$, while keeping $\lambda_x = 0.1$, using the aug-cc-pVDZ basis set.
}

\label{fig:omega_variation}
\end{figure}

The results depicted in Fig.~\ref{fig:omega_variation} using aug-cc-pvDZ~\cite{kendall1992electron} basis set illustrate the correlation energies for the CCSD(2, 1) and CCSD(2, 2) methods across various values of the parameter $\Omega_{cav}$. The correlation energy for the standard CCSD method remains constant at $-0.8275058$, while the CCSD(2, 0) correlation energy remains steady at $-0.8464789$, as they are independent of the $\Omega_{cav}$ value. The SCF energy for purely electronic degrees of freedom is $-245.844381$, and for the QED-HF case, it is $-245.7400942$. The QED-SCF energy also does not vary with the employed $\Omega_{cav}$.

The correlation energy for the CCSD(2, 1) method decreases with increasing $\Omega_{cav}$, starting from $-0.8464789$ at $\Omega_{cav} = 0$ and decreasing to $-0.8750895$ at $\Omega_{cav} = 12$. This indicates that the CCSD(2, 1) method is responsive to changes in $\Omega_{cav}$, capturing more electron-photon interaction as $\Omega_{cav}$ increases. Similarly, the correlation energy for the CCSD(2, 2) method decreases with increasing $\Omega_{cav}$, starting from $-0.8464789$ at $\Omega_{cav} = 0$ and decreasing to $-0.8777892$ at $\Omega_{cav} = 12$. The CCSD(2, 2) method exhibits the most significant variation among the methods, suggesting that it comprehensively captures the electron-photon interaction effects.

Both the CCSD(2, 1) and CCSD(2, 2) methods display a decreasing trend in correlation energy with increasing $\Omega_{cav}$, indicating their ability to better capture the dynamics introduced by varying the electron-photon coupling strength. The larger decrease in correlation energy for the CCSD(2, 2) method compared to CCSD(2, 1) suggests that higher-level approximations within the CCSD framework are more effective in capturing the effects of increasing electron-photon interactions.

These findings imply that as the parameter $\Omega_{cav}$ increases, the system's correlation energy decreases, possibly due to enhanced electron-photon interactions. This emphasizes the importance of employing appropriate quantum electrodynamic coupled-cluster methods to accurately model such systems. For precise modeling of systems where electron-photon interactions are significant, higher-level QED-CCSD methods (such as CCSD(2, 1) and CCSD(2, 2)) should be preferred over standard electronic CCSD and CCSD(2, 0) methods. The data provided demonstrates that higher-level QED-CCSD methods, particularly CCSD(2, 1) and CCSD(2, 2), are sensitive to changes in the electron-photon coupling parameter $\Omega_{cav}$, capturing more nuanced interactions and providing more accurate correlation energies, and thereby changes the electronic ground state.

\section{Conclusion}\label{conclusion}
In summary, our implementation of the quantum electrodynamics coupled-cluster method with single and double excitations (QED-CCSD) within the ExaChem program package marks a significant advancement in computational quantum chemistry. By leveraging the Tensor Algebra for Many-body Methods (TAMM) infrastructure, our framework offers a scalable and efficient solution for modeling both electronic and bosonic amplitudes, encompassing individual and mixed excitation processes.

This innovative computational framework extends beyond the traditional CCSD method, incorporating the intricate interplay between electronic and bosonic degrees of freedom. Through rigorous theoretical foundations, detailed algorithmic descriptions, and comprehensive numerical benchmarks, we have showcased the capability of our approach in accurately describing quantum phenomena, particularly in systems where electron-photon interactions play a significant role.

Our modeling of electron-photon interactions within an optical cavity, using the Pauli-Fierz Hamiltonian within the dipole approximation, adds a new dimension to quantum chemistry simulations. By integrating QED effects into the CCSD framework, we have created a more versatile and accurate model for simulating complex quantum systems. This advancement not only enhances our understanding of fundamental physical processes but also opens new avenues for predicting and manipulating various phenomena with unprecedented precision.
\section*{ACKNOWLEDGMENTS}
The authors acknowledge support from the Center for Many-Body Methods, Spectroscopies, and Dynamics for Molecular Polaritonic Systems (MAPOL) under FWP79715. This support is part of the Computational Chemical Sciences (CCS) program funded by the U.S. Department of Energy, Office of Science, Office of Basic Energy Sciences, Division of Chemical Sciences, Geosciences, and Biosciences at Pacific Northwest National Laboratory (PNNL). PNNL is a multi-program national laboratory operated by Battelle Memorial Institute for the United States Department of Energy under DOE contract number DE-AC05-76RL01830.


\begin{thebibliography}{0}%
\makeatletter
\providecommand \@ifxundefined [1]{%
 \@ifx{#1\undefined}
}%
\providecommand \@ifnum [1]{%
 \ifnum #1\expandafter \@firstoftwo
 \else \expandafter \@secondoftwo
 \fi
}%
\providecommand \@ifx [1]{%
 \ifx #1\expandafter \@firstoftwo
 \else \expandafter \@secondoftwo
 \fi
}%
\providecommand \natexlab [1]{#1}%
\providecommand \enquote  [1]{``#1''}%
\providecommand \bibnamefont  [1]{#1}%
\providecommand \bibfnamefont [1]{#1}%
\providecommand \citenamefont [1]{#1}%
\providecommand \href@noop [0]{\@secondoftwo}%
\providecommand \href [0]{\begingroup \@sanitize@url \@href}%
\providecommand \@href[1]{\@@startlink{#1}\@@href}%
\providecommand \@@href[1]{\endgroup#1\@@endlink}%
\providecommand \@sanitize@url [0]{\catcode `\\12\catcode `\$12\catcode
  `\&12\catcode `\#12\catcode `\^12\catcode `\_12\catcode `\%12\relax}%
\providecommand \@@startlink[1]{}%
\providecommand \@@endlink[0]{}%
\providecommand \url  [0]{\begingroup\@sanitize@url \@url }%
\providecommand \@url [1]{\endgroup\@href {#1}{\urlprefix }}%
\providecommand \urlprefix  [0]{URL }%
\providecommand \Eprint [0]{\href }%
\providecommand \doibase [0]{https://doi.org/}%
\providecommand \selectlanguage [0]{\@gobble}%
\providecommand \bibinfo  [0]{\@secondoftwo}%
\providecommand \bibfield  [0]{\@secondoftwo}%
\providecommand \translation [1]{[#1]}%
\providecommand \BibitemOpen [0]{}%
\providecommand \bibitemStop [0]{}%
\providecommand \bibitemNoStop [0]{.\EOS\space}%
\providecommand \EOS [0]{\spacefactor3000\relax}%
\providecommand \BibitemShut  [1]{\csname bibitem#1\endcsname}%
\let\auto@bib@innerbib\@empty
\end{thebibliography}%


\begin{thebibliography}{75}%
\makeatletter
\providecommand \@ifxundefined [1]{%
 \@ifx{#1\undefined}
}%
\providecommand \@ifnum [1]{%
 \ifnum #1\expandafter \@firstoftwo
 \else \expandafter \@secondoftwo
 \fi
}%
\providecommand \@ifx [1]{%
 \ifx #1\expandafter \@firstoftwo
 \else \expandafter \@secondoftwo
 \fi
}%
\providecommand \natexlab [1]{#1}%
\providecommand \enquote  [1]{``#1''}%
\providecommand \bibnamefont  [1]{#1}%
\providecommand \bibfnamefont [1]{#1}%
\providecommand \citenamefont [1]{#1}%
\providecommand \href@noop [0]{\@secondoftwo}%
\providecommand \href [0]{\begingroup \@sanitize@url \@href}%
\providecommand \@href[1]{\@@startlink{#1}\@@href}%
\providecommand \@@href[1]{\endgroup#1\@@endlink}%
\providecommand \@sanitize@url [0]{\catcode `\\12\catcode `\$12\catcode
  `\&12\catcode `\#12\catcode `\^12\catcode `\_12\catcode `\%12\relax}%
\providecommand \@@startlink[1]{}%
\providecommand \@@endlink[0]{}%
\providecommand \url  [0]{\begingroup\@sanitize@url \@url }%
\providecommand \@url [1]{\endgroup\@href {#1}{\urlprefix }}%
\providecommand \urlprefix  [0]{URL }%
\providecommand \Eprint [0]{\href }%
\providecommand \doibase [0]{https://doi.org/}%
\providecommand \selectlanguage [0]{\@gobble}%
\providecommand \bibinfo  [0]{\@secondoftwo}%
\providecommand \bibfield  [0]{\@secondoftwo}%
\providecommand \translation [1]{[#1]}%
\providecommand \BibitemOpen [0]{}%
\providecommand \bibitemStop [0]{}%
\providecommand \bibitemNoStop [0]{.\EOS\space}%
\providecommand \EOS [0]{\spacefactor3000\relax}%
\providecommand \BibitemShut  [1]{\csname bibitem#1\endcsname}%
\let\auto@bib@innerbib\@empty
\bibitem [{\citenamefont {Helgaker}\ \emph {et~al.}(2013)\citenamefont
  {Helgaker}, \citenamefont {Jorgensen},\ and\ \citenamefont
  {Olsen}}]{helgaker2013molecular}%
  \BibitemOpen
  \bibfield  {author} {\bibinfo {author} {\bibfnamefont {T.}~\bibnamefont
  {Helgaker}}, \bibinfo {author} {\bibfnamefont {P.}~\bibnamefont
  {Jorgensen}},\ and\ \bibinfo {author} {\bibfnamefont {J.}~\bibnamefont
  {Olsen}},\ }\href@noop {} {\emph {\bibinfo {title} {Molecular
  electronic-structure theory}}}\ (\bibinfo  {publisher} {John Wiley \& Sons},\
  \bibinfo {year} {2013})\BibitemShut {NoStop}%
\bibitem [{\citenamefont {Meyer}\ \emph {et~al.}(2009)\citenamefont {Meyer},
  \citenamefont {Gatti},\ and\ \citenamefont
  {Worth}}]{meyer2009multidimensional}%
  \BibitemOpen
  \bibfield  {author} {\bibinfo {author} {\bibfnamefont {H.-D.}\ \bibnamefont
  {Meyer}}, \bibinfo {author} {\bibfnamefont {F.}~\bibnamefont {Gatti}},\ and\
  \bibinfo {author} {\bibfnamefont {G.~A.}\ \bibnamefont {Worth}},\ }\href@noop
  {} {\emph {\bibinfo {title} {Multidimensional quantum dynamics: MCTDH theory
  and applications}}}\ (\bibinfo  {publisher} {John Wiley \& Sons},\ \bibinfo
  {year} {2009})\BibitemShut {NoStop}%
\bibitem [{\citenamefont {Grant}(2007)}]{grant2007relativistic}%
  \BibitemOpen
  \bibfield  {author} {\bibinfo {author} {\bibfnamefont {I.~P.}\ \bibnamefont
  {Grant}},\ }\href@noop {} {\emph {\bibinfo {title} {Relativistic quantum
  theory of atoms and molecules: theory and computation}}}\ (\bibinfo
  {publisher} {Springer},\ \bibinfo {year} {2007})\BibitemShut {NoStop}%
\bibitem [{\citenamefont {Dyall}\ and\ \citenamefont
  {F{\ae}gri~Jr}(2007)}]{dyall2007introduction}%
  \BibitemOpen
  \bibfield  {author} {\bibinfo {author} {\bibfnamefont {K.~G.}\ \bibnamefont
  {Dyall}}\ and\ \bibinfo {author} {\bibfnamefont {K.}~\bibnamefont
  {F{\ae}gri~Jr}},\ }\href@noop {} {\emph {\bibinfo {title} {Introduction to
  relativistic quantum chemistry}}}\ (\bibinfo  {publisher} {Oxford University
  Press},\ \bibinfo {year} {2007})\BibitemShut {NoStop}%
\bibitem [{\citenamefont {Chang}\ \emph {et~al.}(2023)\citenamefont {Chang},
  \citenamefont {Deringer}, \citenamefont {Katti}, \citenamefont
  {Van~Speybroeck},\ and\ \citenamefont {Wolverton}}]{chang2023simulations}%
  \BibitemOpen
  \bibfield  {author} {\bibinfo {author} {\bibfnamefont {C.}~\bibnamefont
  {Chang}}, \bibinfo {author} {\bibfnamefont {V.~L.}\ \bibnamefont {Deringer}},
  \bibinfo {author} {\bibfnamefont {K.~S.}\ \bibnamefont {Katti}}, \bibinfo
  {author} {\bibfnamefont {V.}~\bibnamefont {Van~Speybroeck}},\ and\ \bibinfo
  {author} {\bibfnamefont {C.~M.}\ \bibnamefont {Wolverton}},\ }\bibfield
  {title} {\bibinfo {title} {Simulations in the era of exascale computing},\
  }\href {https://doi.org/https://doi.org/10.1038/s41578-023-00540-6}
  {\bibfield  {journal} {\bibinfo  {journal} {Nature Reviews Materials}\
  }\textbf {\bibinfo {volume} {8}},\ \bibinfo {pages} {309} (\bibinfo {year}
  {2023})}\BibitemShut {NoStop}%
\bibitem [{\citenamefont {Kowalski}\ \emph {et~al.}(2021)\citenamefont
  {Kowalski}, \citenamefont {Bair}, \citenamefont {Bauman}, \citenamefont
  {Boschen}, \citenamefont {Bylaska}, \citenamefont {Daily}, \citenamefont
  {de~Jong}, \citenamefont {Dunning~Jr}, \citenamefont {Govind}, \citenamefont
  {Harrison} \emph {et~al.}}]{kowalski2021nwchem}%
  \BibitemOpen
  \bibfield  {author} {\bibinfo {author} {\bibfnamefont {K.}~\bibnamefont
  {Kowalski}}, \bibinfo {author} {\bibfnamefont {R.}~\bibnamefont {Bair}},
  \bibinfo {author} {\bibfnamefont {N.~P.}\ \bibnamefont {Bauman}}, \bibinfo
  {author} {\bibfnamefont {J.~S.}\ \bibnamefont {Boschen}}, \bibinfo {author}
  {\bibfnamefont {E.~J.}\ \bibnamefont {Bylaska}}, \bibinfo {author}
  {\bibfnamefont {J.}~\bibnamefont {Daily}}, \bibinfo {author} {\bibfnamefont
  {W.~A.}\ \bibnamefont {de~Jong}}, \bibinfo {author} {\bibfnamefont
  {T.}~\bibnamefont {Dunning~Jr}}, \bibinfo {author} {\bibfnamefont
  {N.}~\bibnamefont {Govind}}, \bibinfo {author} {\bibfnamefont {R.~J.}\
  \bibnamefont {Harrison}}, \emph {et~al.},\ }\bibfield  {title} {\bibinfo
  {title} {From nwchem to nwchemex: Evolving with the computational chemistry
  landscape},\ }\href
  {https://doi.org/https://doi.org/10.1021/acs.chemrev.0c00998} {\bibfield
  {journal} {\bibinfo  {journal} {Chemical reviews}\ }\textbf {\bibinfo
  {volume} {121}},\ \bibinfo {pages} {4962} (\bibinfo {year}
  {2021})}\BibitemShut {NoStop}%
\bibitem [{\citenamefont {Ruggenthaler}\ \emph {et~al.}(2018)\citenamefont
  {Ruggenthaler}, \citenamefont {Tancogne-Dejean}, \citenamefont {Flick},
  \citenamefont {Appel},\ and\ \citenamefont
  {Rubio}}]{ruggenthaler2018quantum}%
  \BibitemOpen
  \bibfield  {author} {\bibinfo {author} {\bibfnamefont {M.}~\bibnamefont
  {Ruggenthaler}}, \bibinfo {author} {\bibfnamefont {N.}~\bibnamefont
  {Tancogne-Dejean}}, \bibinfo {author} {\bibfnamefont {J.}~\bibnamefont
  {Flick}}, \bibinfo {author} {\bibfnamefont {H.}~\bibnamefont {Appel}},\ and\
  \bibinfo {author} {\bibfnamefont {A.}~\bibnamefont {Rubio}},\ }\bibfield
  {title} {\bibinfo {title} {From a quantum-electrodynamical light--matter
  description to novel spectroscopies},\ }\href
  {https://doi.org/https://doi.org/10.1038/s41570-018-0118} {\bibfield
  {journal} {\bibinfo  {journal} {Nature Reviews Chemistry}\ }\textbf {\bibinfo
  {volume} {2}},\ \bibinfo {pages} {1} (\bibinfo {year} {2018})}\BibitemShut
  {NoStop}%
\bibitem [{\citenamefont {H{\"u}bener}\ \emph {et~al.}(2017)\citenamefont
  {H{\"u}bener}, \citenamefont {Sentef}, \citenamefont {De~Giovannini},
  \citenamefont {Kemper},\ and\ \citenamefont {Rubio}}]{hubener2017creating}%
  \BibitemOpen
  \bibfield  {author} {\bibinfo {author} {\bibfnamefont {H.}~\bibnamefont
  {H{\"u}bener}}, \bibinfo {author} {\bibfnamefont {M.~A.}\ \bibnamefont
  {Sentef}}, \bibinfo {author} {\bibfnamefont {U.}~\bibnamefont
  {De~Giovannini}}, \bibinfo {author} {\bibfnamefont {A.~F.}\ \bibnamefont
  {Kemper}},\ and\ \bibinfo {author} {\bibfnamefont {A.}~\bibnamefont
  {Rubio}},\ }\bibfield  {title} {\bibinfo {title} {Creating stable
  floquet--weyl semimetals by laser-driving of 3d dirac materials},\ }\href
  {https://doi.org/https://doi.org/10.1038/ncomms13940} {\bibfield  {journal}
  {\bibinfo  {journal} {Nature communications}\ }\textbf {\bibinfo {volume}
  {8}},\ \bibinfo {pages} {13940} (\bibinfo {year} {2017})}\BibitemShut
  {NoStop}%
\bibitem [{\citenamefont {Byrnes}\ \emph {et~al.}(2014)\citenamefont {Byrnes},
  \citenamefont {Kim},\ and\ \citenamefont {Yamamoto}}]{byrnes2014exciton}%
  \BibitemOpen
  \bibfield  {author} {\bibinfo {author} {\bibfnamefont {T.}~\bibnamefont
  {Byrnes}}, \bibinfo {author} {\bibfnamefont {N.~Y.}\ \bibnamefont {Kim}},\
  and\ \bibinfo {author} {\bibfnamefont {Y.}~\bibnamefont {Yamamoto}},\
  }\bibfield  {title} {\bibinfo {title} {Exciton--polariton condensates},\
  }\href {https://doi.org/https://doi.org/10.1038/nphys3143} {\bibfield
  {journal} {\bibinfo  {journal} {Nature Physics}\ }\textbf {\bibinfo {volume}
  {10}},\ \bibinfo {pages} {803} (\bibinfo {year} {2014})}\BibitemShut
  {NoStop}%
\bibitem [{\citenamefont {Basov}\ \emph {et~al.}(2016)\citenamefont {Basov},
  \citenamefont {Fogler},\ and\ \citenamefont {Garc{\'\i}a~de
  Abajo}}]{basov2016polaritons}%
  \BibitemOpen
  \bibfield  {author} {\bibinfo {author} {\bibfnamefont {D.}~\bibnamefont
  {Basov}}, \bibinfo {author} {\bibfnamefont {M.}~\bibnamefont {Fogler}},\ and\
  \bibinfo {author} {\bibfnamefont {F.}~\bibnamefont {Garc{\'\i}a~de Abajo}},\
  }\bibfield  {title} {\bibinfo {title} {Polaritons in van der waals
  materials},\ }\href {https://doi.org/10.1126/science.aag1992} {\bibfield
  {journal} {\bibinfo  {journal} {Science}\ }\textbf {\bibinfo {volume}
  {354}},\ \bibinfo {pages} {aag1992} (\bibinfo {year} {2016})}\BibitemShut
  {NoStop}%
\bibitem [{\citenamefont {Latini}\ \emph {et~al.}(2019)\citenamefont {Latini},
  \citenamefont {Ronca}, \citenamefont {De~Giovannini}, \citenamefont
  {H\"{u}bener},\ and\ \citenamefont {Rubio}}]{latini2019cavity}%
  \BibitemOpen
  \bibfield  {author} {\bibinfo {author} {\bibfnamefont {S.}~\bibnamefont
  {Latini}}, \bibinfo {author} {\bibfnamefont {E.}~\bibnamefont {Ronca}},
  \bibinfo {author} {\bibfnamefont {U.}~\bibnamefont {De~Giovannini}}, \bibinfo
  {author} {\bibfnamefont {H.}~\bibnamefont {H\"{u}bener}},\ and\ \bibinfo
  {author} {\bibfnamefont {A.}~\bibnamefont {Rubio}},\ }\bibfield  {title}
  {\bibinfo {title} {Cavity control of excitons in two-dimensional materials},\
  }\href@noop {} {\bibfield  {journal} {\bibinfo  {journal} {Nano letters}\
  }\textbf {\bibinfo {volume} {19}},\ \bibinfo {pages} {3473} (\bibinfo {year}
  {2019})}\BibitemShut {NoStop}%
\bibitem [{\citenamefont {Thomas}\ \emph {et~al.}(2016)\citenamefont {Thomas},
  \citenamefont {George}, \citenamefont {Shalabney}, \citenamefont {Dryzhakov},
  \citenamefont {Varma}, \citenamefont {Moran}, \citenamefont {Chervy},
  \citenamefont {Zhong}, \citenamefont {Devaux}, \citenamefont {Genet},
  \citenamefont {Hutchison},\ and\ \citenamefont {Ebbesen}}]{Thomas}%
  \BibitemOpen
  \bibfield  {author} {\bibinfo {author} {\bibfnamefont {A.}~\bibnamefont
  {Thomas}}, \bibinfo {author} {\bibfnamefont {J.}~\bibnamefont {George}},
  \bibinfo {author} {\bibfnamefont {A.}~\bibnamefont {Shalabney}}, \bibinfo
  {author} {\bibfnamefont {M.}~\bibnamefont {Dryzhakov}}, \bibinfo {author}
  {\bibfnamefont {S.~J.}\ \bibnamefont {Varma}}, \bibinfo {author}
  {\bibfnamefont {J.}~\bibnamefont {Moran}}, \bibinfo {author} {\bibfnamefont
  {T.}~\bibnamefont {Chervy}}, \bibinfo {author} {\bibfnamefont
  {X.}~\bibnamefont {Zhong}}, \bibinfo {author} {\bibfnamefont
  {E.}~\bibnamefont {Devaux}}, \bibinfo {author} {\bibfnamefont
  {C.}~\bibnamefont {Genet}}, \bibinfo {author} {\bibfnamefont {J.~A.}\
  \bibnamefont {Hutchison}},\ and\ \bibinfo {author} {\bibfnamefont {T.~W.}\
  \bibnamefont {Ebbesen}},\ }\bibfield  {title} {\bibinfo {title} {Ground-state
  chemical reactivity under vibrational coupling to the vacuum electromagnetic
  field},\ }\href {https://doi.org/https://doi.org/10.1002/anie.201605504}
  {\bibfield  {journal} {\bibinfo  {journal} {Angewandte Chemie International
  Edition}\ }\textbf {\bibinfo {volume} {55}},\ \bibinfo {pages} {11462}
  (\bibinfo {year} {2016})}\BibitemShut {NoStop}%
\bibitem [{\citenamefont {Thomas}\ \emph {et~al.}(2019)\citenamefont {Thomas},
  \citenamefont {Lethuillier-Karl}, \citenamefont {Nagarajan}, \citenamefont
  {Vergauwe}, \citenamefont {George}, \citenamefont {Chervy}, \citenamefont
  {Shalabney}, \citenamefont {Devaux}, \citenamefont {Genet}, \citenamefont
  {Moran} \emph {et~al.}}]{thomas2}%
  \BibitemOpen
  \bibfield  {author} {\bibinfo {author} {\bibfnamefont {A.}~\bibnamefont
  {Thomas}}, \bibinfo {author} {\bibfnamefont {L.}~\bibnamefont
  {Lethuillier-Karl}}, \bibinfo {author} {\bibfnamefont {K.}~\bibnamefont
  {Nagarajan}}, \bibinfo {author} {\bibfnamefont {R.~M.}\ \bibnamefont
  {Vergauwe}}, \bibinfo {author} {\bibfnamefont {J.}~\bibnamefont {George}},
  \bibinfo {author} {\bibfnamefont {T.}~\bibnamefont {Chervy}}, \bibinfo
  {author} {\bibfnamefont {A.}~\bibnamefont {Shalabney}}, \bibinfo {author}
  {\bibfnamefont {E.}~\bibnamefont {Devaux}}, \bibinfo {author} {\bibfnamefont
  {C.}~\bibnamefont {Genet}}, \bibinfo {author} {\bibfnamefont
  {J.}~\bibnamefont {Moran}}, \emph {et~al.},\ }\bibfield  {title} {\bibinfo
  {title} {Tilting a ground-state reactivity landscape by vibrational strong
  coupling},\ }\href@noop {} {\bibfield  {journal} {\bibinfo  {journal}
  {Science}\ }\textbf {\bibinfo {volume} {363}},\ \bibinfo {pages} {615}
  (\bibinfo {year} {2019})}\BibitemShut {NoStop}%
\bibitem [{\citenamefont {Sidler}\ \emph {et~al.}(2022)\citenamefont {Sidler},
  \citenamefont {Ruggenthaler}, \citenamefont {Sch\"{a}fer}, \citenamefont
  {Ronca},\ and\ \citenamefont {Rubio}}]{rubio_review}%
  \BibitemOpen
  \bibfield  {author} {\bibinfo {author} {\bibfnamefont {D.}~\bibnamefont
  {Sidler}}, \bibinfo {author} {\bibfnamefont {M.}~\bibnamefont
  {Ruggenthaler}}, \bibinfo {author} {\bibfnamefont {C.}~\bibnamefont
  {Sch\"{a}fer}}, \bibinfo {author} {\bibfnamefont {E.}~\bibnamefont {Ronca}},\
  and\ \bibinfo {author} {\bibfnamefont {A.}~\bibnamefont {Rubio}},\ }\bibfield
   {title} {\bibinfo {title} {{A perspective on ab initio modeling of
  polaritonic chemistry: The role of non-equilibrium effects and quantum
  collectivity}},\ }\href {https://doi.org/10.1063/5.0094956} {\bibfield
  {journal} {\bibinfo  {journal} {The Journal of Chemical Physics}\ }\textbf
  {\bibinfo {volume} {156}},\ \bibinfo {pages} {230901} (\bibinfo {year}
  {2022})}\BibitemShut {NoStop}%
\bibitem [{\citenamefont {Galego}\ \emph {et~al.}(2019)\citenamefont {Galego},
  \citenamefont {Climent}, \citenamefont {Garcia-Vidal},\ and\ \citenamefont
  {Feist}}]{feist1}%
  \BibitemOpen
  \bibfield  {author} {\bibinfo {author} {\bibfnamefont {J.}~\bibnamefont
  {Galego}}, \bibinfo {author} {\bibfnamefont {C.}~\bibnamefont {Climent}},
  \bibinfo {author} {\bibfnamefont {F.~J.}\ \bibnamefont {Garcia-Vidal}},\ and\
  \bibinfo {author} {\bibfnamefont {J.}~\bibnamefont {Feist}},\ }\bibfield
  {title} {\bibinfo {title} {Cavity casimir-polder forces and their effects in
  ground-state chemical reactivity},\ }\href
  {https://doi.org/10.1103/PhysRevX.9.021057} {\bibfield  {journal} {\bibinfo
  {journal} {Phys. Rev. X}\ }\textbf {\bibinfo {volume} {9}},\ \bibinfo {pages}
  {021057} (\bibinfo {year} {2019})}\BibitemShut {NoStop}%
\bibitem [{\citenamefont {Galego}\ \emph {et~al.}(2015)\citenamefont {Galego},
  \citenamefont {Garcia-Vidal},\ and\ \citenamefont {Feist}}]{feist2}%
  \BibitemOpen
  \bibfield  {author} {\bibinfo {author} {\bibfnamefont {J.}~\bibnamefont
  {Galego}}, \bibinfo {author} {\bibfnamefont {F.~J.}\ \bibnamefont
  {Garcia-Vidal}},\ and\ \bibinfo {author} {\bibfnamefont {J.}~\bibnamefont
  {Feist}},\ }\bibfield  {title} {\bibinfo {title} {Cavity-induced
  modifications of molecular structure in the strong-coupling regime},\ }\href
  {https://doi.org/10.1103/PhysRevX.5.041022} {\bibfield  {journal} {\bibinfo
  {journal} {Phys. Rev. X}\ }\textbf {\bibinfo {volume} {5}},\ \bibinfo {pages}
  {041022} (\bibinfo {year} {2015})}\BibitemShut {NoStop}%
\bibitem [{\citenamefont {Walther}\ \emph {et~al.}(2006)\citenamefont
  {Walther}, \citenamefont {Varcoe}, \citenamefont {Englert},\ and\
  \citenamefont {Becker}}]{Walther_2006}%
  \BibitemOpen
  \bibfield  {author} {\bibinfo {author} {\bibfnamefont {H.}~\bibnamefont
  {Walther}}, \bibinfo {author} {\bibfnamefont {B.~T.~H.}\ \bibnamefont
  {Varcoe}}, \bibinfo {author} {\bibfnamefont {B.-G.}\ \bibnamefont
  {Englert}},\ and\ \bibinfo {author} {\bibfnamefont {T.}~\bibnamefont
  {Becker}},\ }\bibfield  {title} {\bibinfo {title} {Cavity quantum
  electrodynamics},\ }\href {https://doi.org/10.1088/0034-4885/69/5/R02}
  {\bibfield  {journal} {\bibinfo  {journal} {Reports on Progress in Physics}\
  }\textbf {\bibinfo {volume} {69}},\ \bibinfo {pages} {1325} (\bibinfo {year}
  {2006})}\BibitemShut {NoStop}%
\bibitem [{\citenamefont {Ribeiro}\ \emph {et~al.}(2018)\citenamefont
  {Ribeiro}, \citenamefont {Mart{\'\i}nez-Mart{\'\i}nez}, \citenamefont {Du},
  \citenamefont {Campos-Gonzalez-Angulo},\ and\ \citenamefont
  {Yuen-Zhou}}]{ribeiro2018polariton}%
  \BibitemOpen
  \bibfield  {author} {\bibinfo {author} {\bibfnamefont {R.~F.}\ \bibnamefont
  {Ribeiro}}, \bibinfo {author} {\bibfnamefont {L.~A.}\ \bibnamefont
  {Mart{\'\i}nez-Mart{\'\i}nez}}, \bibinfo {author} {\bibfnamefont
  {M.}~\bibnamefont {Du}}, \bibinfo {author} {\bibfnamefont {J.}~\bibnamefont
  {Campos-Gonzalez-Angulo}},\ and\ \bibinfo {author} {\bibfnamefont
  {J.}~\bibnamefont {Yuen-Zhou}},\ }\bibfield  {title} {\bibinfo {title}
  {Polariton chemistry: controlling molecular dynamics with optical cavities},\
  }\href@noop {} {\bibfield  {journal} {\bibinfo  {journal} {Chemical science}\
  }\textbf {\bibinfo {volume} {9}},\ \bibinfo {pages} {6325} (\bibinfo {year}
  {2018})}\BibitemShut {NoStop}%
\bibitem [{\citenamefont {Campos-Gonzalez-Angulo}\ \emph
  {et~al.}(2019)\citenamefont {Campos-Gonzalez-Angulo}, \citenamefont
  {Ribeiro},\ and\ \citenamefont {Yuen-Zhou}}]{campos2019resonant}%
  \BibitemOpen
  \bibfield  {author} {\bibinfo {author} {\bibfnamefont {J.~A.}\ \bibnamefont
  {Campos-Gonzalez-Angulo}}, \bibinfo {author} {\bibfnamefont {R.~F.}\
  \bibnamefont {Ribeiro}},\ and\ \bibinfo {author} {\bibfnamefont
  {J.}~\bibnamefont {Yuen-Zhou}},\ }\bibfield  {title} {\bibinfo {title}
  {Resonant catalysis of thermally activated chemical reactions with
  vibrational polaritons},\ }\href@noop {} {\bibfield  {journal} {\bibinfo
  {journal} {Nature communications}\ }\textbf {\bibinfo {volume} {10}},\
  \bibinfo {pages} {4685} (\bibinfo {year} {2019})}\BibitemShut {NoStop}%
\bibitem [{\citenamefont {Gallego-Valencia}\ \emph {et~al.}(2024)\citenamefont
  {Gallego-Valencia}, \citenamefont {Mewes}, \citenamefont {Feist},\ and\
  \citenamefont {Sanz-Vicario}}]{gallego2024coherent}%
  \BibitemOpen
  \bibfield  {author} {\bibinfo {author} {\bibfnamefont {D.}~\bibnamefont
  {Gallego-Valencia}}, \bibinfo {author} {\bibfnamefont {L.}~\bibnamefont
  {Mewes}}, \bibinfo {author} {\bibfnamefont {J.}~\bibnamefont {Feist}},\ and\
  \bibinfo {author} {\bibfnamefont {J.~L.}\ \bibnamefont {Sanz-Vicario}},\
  }\bibfield  {title} {\bibinfo {title} {Coherent multidimensional spectroscopy
  in polariton systems},\ }\href@noop {} {\bibfield  {journal} {\bibinfo
  {journal} {arXiv preprint arXiv:2403.04734}\ } (\bibinfo {year}
  {2024})}\BibitemShut {NoStop}%
\bibitem [{\citenamefont {Ruggenthaler}\ \emph {et~al.}(2023)\citenamefont
  {Ruggenthaler}, \citenamefont {Sidler},\ and\ \citenamefont
  {Rubio}}]{rubio_chemrev}%
  \BibitemOpen
  \bibfield  {author} {\bibinfo {author} {\bibfnamefont {M.}~\bibnamefont
  {Ruggenthaler}}, \bibinfo {author} {\bibfnamefont {D.}~\bibnamefont
  {Sidler}},\ and\ \bibinfo {author} {\bibfnamefont {A.}~\bibnamefont
  {Rubio}},\ }\bibfield  {title} {\bibinfo {title} {Understanding polaritonic
  chemistry from ab initio quantum electrodynamics},\ }\href
  {https://doi.org/10.1021/acs.chemrev.2c00788} {\bibfield  {journal} {\bibinfo
   {journal} {Chemical Reviews}\ }\textbf {\bibinfo {volume} {123}},\ \bibinfo
  {pages} {11191} (\bibinfo {year} {2023})}\BibitemShut {NoStop}%
\bibitem [{\citenamefont {Ruggenthaler}\ \emph {et~al.}(2011)\citenamefont
  {Ruggenthaler}, \citenamefont {Mackenroth},\ and\ \citenamefont
  {Bauer}}]{PhysRevA.84.042107}%
  \BibitemOpen
  \bibfield  {author} {\bibinfo {author} {\bibfnamefont {M.}~\bibnamefont
  {Ruggenthaler}}, \bibinfo {author} {\bibfnamefont {F.}~\bibnamefont
  {Mackenroth}},\ and\ \bibinfo {author} {\bibfnamefont {D.}~\bibnamefont
  {Bauer}},\ }\bibfield  {title} {\bibinfo {title} {Time-dependent kohn-sham
  approach to quantum electrodynamics},\ }\href
  {https://doi.org/10.1103/PhysRevA.84.042107} {\bibfield  {journal} {\bibinfo
  {journal} {Phys. Rev. A}\ }\textbf {\bibinfo {volume} {84}},\ \bibinfo
  {pages} {042107} (\bibinfo {year} {2011})}\BibitemShut {NoStop}%
\bibitem [{\citenamefont {Ruggenthaler}\ \emph {et~al.}(2014)\citenamefont
  {Ruggenthaler}, \citenamefont {Flick}, \citenamefont {Pellegrini},
  \citenamefont {Appel}, \citenamefont {Tokatly},\ and\ \citenamefont
  {Rubio}}]{PhysRevA.90.012508}%
  \BibitemOpen
  \bibfield  {author} {\bibinfo {author} {\bibfnamefont {M.}~\bibnamefont
  {Ruggenthaler}}, \bibinfo {author} {\bibfnamefont {J.}~\bibnamefont {Flick}},
  \bibinfo {author} {\bibfnamefont {C.}~\bibnamefont {Pellegrini}}, \bibinfo
  {author} {\bibfnamefont {H.}~\bibnamefont {Appel}}, \bibinfo {author}
  {\bibfnamefont {I.~V.}\ \bibnamefont {Tokatly}},\ and\ \bibinfo {author}
  {\bibfnamefont {A.}~\bibnamefont {Rubio}},\ }\bibfield  {title} {\bibinfo
  {title} {Quantum-electrodynamical density-functional theory: Bridging quantum
  optics and electronic-structure theory},\ }\href
  {https://doi.org/10.1103/PhysRevA.90.012508} {\bibfield  {journal} {\bibinfo
  {journal} {Phys. Rev. A}\ }\textbf {\bibinfo {volume} {90}},\ \bibinfo
  {pages} {012508} (\bibinfo {year} {2014})}\BibitemShut {NoStop}%
\bibitem [{\citenamefont {Pellegrini}\ \emph {et~al.}(2015)\citenamefont
  {Pellegrini}, \citenamefont {Flick}, \citenamefont {Tokatly}, \citenamefont
  {Appel},\ and\ \citenamefont {Rubio}}]{PhysRevLett.115.093001}%
  \BibitemOpen
  \bibfield  {author} {\bibinfo {author} {\bibfnamefont {C.}~\bibnamefont
  {Pellegrini}}, \bibinfo {author} {\bibfnamefont {J.}~\bibnamefont {Flick}},
  \bibinfo {author} {\bibfnamefont {I.~V.}\ \bibnamefont {Tokatly}}, \bibinfo
  {author} {\bibfnamefont {H.}~\bibnamefont {Appel}},\ and\ \bibinfo {author}
  {\bibfnamefont {A.}~\bibnamefont {Rubio}},\ }\bibfield  {title} {\bibinfo
  {title} {Optimized effective potential for quantum electrodynamical
  time-dependent density functional theory},\ }\href
  {https://doi.org/10.1103/PhysRevLett.115.093001} {\bibfield  {journal}
  {\bibinfo  {journal} {Phys. Rev. Lett.}\ }\textbf {\bibinfo {volume} {115}},\
  \bibinfo {pages} {093001} (\bibinfo {year} {2015})}\BibitemShut {NoStop}%
\bibitem [{\citenamefont {Flick}\ \emph {et~al.}(2018)\citenamefont {Flick},
  \citenamefont {Sch\"{a}fer}, \citenamefont {Ruggenthaler}, \citenamefont
  {Appel},\ and\ \citenamefont {Rubio}}]{doi:10.1021/acsphotonics.7b01279}%
  \BibitemOpen
  \bibfield  {author} {\bibinfo {author} {\bibfnamefont {J.}~\bibnamefont
  {Flick}}, \bibinfo {author} {\bibfnamefont {C.}~\bibnamefont {Sch\"{a}fer}},
  \bibinfo {author} {\bibfnamefont {M.}~\bibnamefont {Ruggenthaler}}, \bibinfo
  {author} {\bibfnamefont {H.}~\bibnamefont {Appel}},\ and\ \bibinfo {author}
  {\bibfnamefont {A.}~\bibnamefont {Rubio}},\ }\bibfield  {title} {\bibinfo
  {title} {Ab initio optimized effective potentials for real molecules in
  optical cavities: Photon contributions to the molecular ground state},\
  }\href {https://doi.org/10.1021/acsphotonics.7b01279} {\bibfield  {journal}
  {\bibinfo  {journal} {ACS Photonics}\ }\textbf {\bibinfo {volume} {5}},\
  \bibinfo {pages} {992} (\bibinfo {year} {2018})}\BibitemShut {NoStop}%
\bibitem [{\citenamefont {Ren\'{e}~Jest\"{a}dt}\ and\ \citenamefont
  {Appel}(2019)}]{doi:10.1080/00018732.2019.1695875}%
  \BibitemOpen
  \bibfield  {author} {\bibinfo {author} {\bibfnamefont {M.~J. T. O. A.~R.}\
  \bibnamefont {Ren\'{e}~Jest\"{a}dt}, \bibfnamefont {Michael~Ruggenthaler}}\
  and\ \bibinfo {author} {\bibfnamefont {H.}~\bibnamefont {Appel}},\ }\bibfield
   {title} {\bibinfo {title} {Light-matter interactions within the
  ehrenfest–maxwell–pauli–kohn–sham framework: fundamentals,
  implementation, and nano-optical applications},\ }\href
  {https://doi.org/10.1080/00018732.2019.1695875} {\bibfield  {journal}
  {\bibinfo  {journal} {Advances in Physics}\ }\textbf {\bibinfo {volume}
  {68}},\ \bibinfo {pages} {225} (\bibinfo {year} {2019})}\BibitemShut
  {NoStop}%
\bibitem [{\citenamefont {Flick}\ and\ \citenamefont
  {Narang}(2020)}]{10.1063/5.0021033}%
  \BibitemOpen
  \bibfield  {author} {\bibinfo {author} {\bibfnamefont {J.}~\bibnamefont
  {Flick}}\ and\ \bibinfo {author} {\bibfnamefont {P.}~\bibnamefont {Narang}},\
  }\bibfield  {title} {\bibinfo {title} {{Ab initio polaritonic
  potential-energy surfaces for excited-state nanophotonics and polaritonic
  chemistry}},\ }\href {https://doi.org/10.1063/5.0021033} {\bibfield
  {journal} {\bibinfo  {journal} {The Journal of Chemical Physics}\ }\textbf
  {\bibinfo {volume} {153}},\ \bibinfo {pages} {094116} (\bibinfo {year}
  {2020})}\BibitemShut {NoStop}%
\bibitem [{\citenamefont {Vu}\ \emph {et~al.}(2022)\citenamefont {Vu},
  \citenamefont {McLeod}, \citenamefont {Hanson},\ and\ \citenamefont
  {DePrince}}]{doi:10.1021/acs.jpca.2c07134}%
  \BibitemOpen
  \bibfield  {author} {\bibinfo {author} {\bibfnamefont {N.}~\bibnamefont
  {Vu}}, \bibinfo {author} {\bibfnamefont {G.~M.}\ \bibnamefont {McLeod}},
  \bibinfo {author} {\bibfnamefont {K.}~\bibnamefont {Hanson}},\ and\ \bibinfo
  {author} {\bibfnamefont {A.~E.~I.}\ \bibnamefont {DePrince}},\ }\bibfield
  {title} {\bibinfo {title} {Enhanced diastereocontrol via strong
  light–matter interactions in an optical cavity},\ }\href
  {https://doi.org/10.1021/acs.jpca.2c07134} {\bibfield  {journal} {\bibinfo
  {journal} {The Journal of Physical Chemistry A}\ }\textbf {\bibinfo {volume}
  {126}},\ \bibinfo {pages} {9303} (\bibinfo {year} {2022})}\BibitemShut
  {NoStop}%
\bibitem [{\citenamefont {Pavo\v{s}evi\'c}\ and\ \citenamefont
  {Rubio}(2022)}]{10.1063/5.0095552}%
  \BibitemOpen
  \bibfield  {author} {\bibinfo {author} {\bibfnamefont {F.}~\bibnamefont
  {Pavo\v{s}evi\'c}}\ and\ \bibinfo {author} {\bibfnamefont {A.}~\bibnamefont
  {Rubio}},\ }\bibfield  {title} {\bibinfo {title} {{Wavefunction embedding for
  molecular polaritons}},\ }\href {https://doi.org/10.1063/5.0095552}
  {\bibfield  {journal} {\bibinfo  {journal} {The Journal of Chemical Physics}\
  }\textbf {\bibinfo {volume} {157}},\ \bibinfo {pages} {094101} (\bibinfo
  {year} {2022})}\BibitemShut {NoStop}%
\bibitem [{\citenamefont {Liebenthal}\ \emph {et~al.}(2023)\citenamefont
  {Liebenthal}, \citenamefont {Vu},\ and\ \citenamefont
  {DePrince}}]{doi:10.1021/acs.jpca.3c01842}%
  \BibitemOpen
  \bibfield  {author} {\bibinfo {author} {\bibfnamefont {M.~D.}\ \bibnamefont
  {Liebenthal}}, \bibinfo {author} {\bibfnamefont {N.}~\bibnamefont {Vu}},\
  and\ \bibinfo {author} {\bibfnamefont {A.~E.~I.}\ \bibnamefont {DePrince}},\
  }\bibfield  {title} {\bibinfo {title} {Assessing the effects of orbital
  relaxation and the coherent-state transformation in quantum electrodynamics
  density functional and coupled-cluster theories},\ }\href
  {https://doi.org/10.1021/acs.jpca.3c01842} {\bibfield  {journal} {\bibinfo
  {journal} {The Journal of Physical Chemistry A}\ }\textbf {\bibinfo {volume}
  {127}},\ \bibinfo {pages} {5264} (\bibinfo {year} {2023})}\BibitemShut
  {NoStop}%
\bibitem [{\citenamefont {Tokatly}(2013)}]{PhysRevLett.110.233001}%
  \BibitemOpen
  \bibfield  {author} {\bibinfo {author} {\bibfnamefont {I.~V.}\ \bibnamefont
  {Tokatly}},\ }\bibfield  {title} {\bibinfo {title} {Time-dependent density
  functional theory for many-electron systems interacting with cavity
  photons},\ }\href {https://doi.org/10.1103/PhysRevLett.110.233001} {\bibfield
   {journal} {\bibinfo  {journal} {Phys. Rev. Lett.}\ }\textbf {\bibinfo
  {volume} {110}},\ \bibinfo {pages} {233001} (\bibinfo {year}
  {2013})}\BibitemShut {NoStop}%
\bibitem [{\citenamefont {Flick}\ \emph {et~al.}(2017)\citenamefont {Flick},
  \citenamefont {Ruggenthaler}, \citenamefont {Appel},\ and\ \citenamefont
  {Rubio}}]{flick2017atoms}%
  \BibitemOpen
  \bibfield  {author} {\bibinfo {author} {\bibfnamefont {J.}~\bibnamefont
  {Flick}}, \bibinfo {author} {\bibfnamefont {M.}~\bibnamefont {Ruggenthaler}},
  \bibinfo {author} {\bibfnamefont {H.}~\bibnamefont {Appel}},\ and\ \bibinfo
  {author} {\bibfnamefont {A.}~\bibnamefont {Rubio}},\ }\bibfield  {title}
  {\bibinfo {title} {Atoms and molecules in cavities, from weak to strong
  coupling in quantum-electrodynamics (qed) chemistry},\ }\href@noop {}
  {\bibfield  {journal} {\bibinfo  {journal} {Proceedings of the National
  Academy of Sciences}\ }\textbf {\bibinfo {volume} {114}},\ \bibinfo {pages}
  {3026} (\bibinfo {year} {2017})}\BibitemShut {NoStop}%
\bibitem [{\citenamefont {Tokatly}(2018)}]{PhysRevB.98.235123}%
  \BibitemOpen
  \bibfield  {author} {\bibinfo {author} {\bibfnamefont {I.~V.}\ \bibnamefont
  {Tokatly}},\ }\bibfield  {title} {\bibinfo {title} {Conserving approximations
  in cavity quantum electrodynamics: Implications for density functional theory
  of electron-photon systems},\ }\href
  {https://doi.org/10.1103/PhysRevB.98.235123} {\bibfield  {journal} {\bibinfo
  {journal} {Phys. Rev. B}\ }\textbf {\bibinfo {volume} {98}},\ \bibinfo
  {pages} {235123} (\bibinfo {year} {2018})}\BibitemShut {NoStop}%
\bibitem [{\citenamefont {Malave}\ \emph {et~al.}(2022)\citenamefont {Malave},
  \citenamefont {Ahrens}, \citenamefont {Pitagora}, \citenamefont {Covington},\
  and\ \citenamefont {Varga}}]{10.1063/5.0123909}%
  \BibitemOpen
  \bibfield  {author} {\bibinfo {author} {\bibfnamefont {J.}~\bibnamefont
  {Malave}}, \bibinfo {author} {\bibfnamefont {A.}~\bibnamefont {Ahrens}},
  \bibinfo {author} {\bibfnamefont {D.}~\bibnamefont {Pitagora}}, \bibinfo
  {author} {\bibfnamefont {C.}~\bibnamefont {Covington}},\ and\ \bibinfo
  {author} {\bibfnamefont {K.}~\bibnamefont {Varga}},\ }\bibfield  {title}
  {\bibinfo {title} {{Real-space, real-time approach to
  quantum-electrodynamical time-dependent density functional theory}},\ }\href
  {https://doi.org/10.1063/5.0123909} {\bibfield  {journal} {\bibinfo
  {journal} {The Journal of Chemical Physics}\ }\textbf {\bibinfo {volume}
  {157}},\ \bibinfo {pages} {194106} (\bibinfo {year} {2022})}\BibitemShut
  {NoStop}%
\bibitem [{\citenamefont {Yang}\ \emph {et~al.}(2021)\citenamefont {Yang},
  \citenamefont {Ou}, \citenamefont {Pei}, \citenamefont {Wang}, \citenamefont
  {Weng}, \citenamefont {Shuai}, \citenamefont {Mullen},\ and\ \citenamefont
  {Shao}}]{10.1063/5.0057542}%
  \BibitemOpen
  \bibfield  {author} {\bibinfo {author} {\bibfnamefont {J.}~\bibnamefont
  {Yang}}, \bibinfo {author} {\bibfnamefont {Q.}~\bibnamefont {Ou}}, \bibinfo
  {author} {\bibfnamefont {Z.}~\bibnamefont {Pei}}, \bibinfo {author}
  {\bibfnamefont {H.}~\bibnamefont {Wang}}, \bibinfo {author} {\bibfnamefont
  {B.}~\bibnamefont {Weng}}, \bibinfo {author} {\bibfnamefont {Z.}~\bibnamefont
  {Shuai}}, \bibinfo {author} {\bibfnamefont {K.}~\bibnamefont {Mullen}},\ and\
  \bibinfo {author} {\bibfnamefont {Y.}~\bibnamefont {Shao}},\ }\bibfield
  {title} {\bibinfo {title} {{Quantum-electrodynamical time-dependent density
  functional theory within Gaussian atomic basis}},\ }\href
  {https://doi.org/10.1063/5.0057542} {\bibfield  {journal} {\bibinfo
  {journal} {The Journal of Chemical Physics}\ }\textbf {\bibinfo {volume}
  {155}},\ \bibinfo {pages} {064107} (\bibinfo {year} {2021})}\BibitemShut
  {NoStop}%
\bibitem [{\citenamefont {Yang}\ \emph {et~al.}(2022)\citenamefont {Yang},
  \citenamefont {Pei}, \citenamefont {Leon}, \citenamefont {Wickizer},
  \citenamefont {Weng}, \citenamefont {Mao}, \citenamefont {Ou},\ and\
  \citenamefont {Shao}}]{10.1063/5.0082386}%
  \BibitemOpen
  \bibfield  {author} {\bibinfo {author} {\bibfnamefont {J.}~\bibnamefont
  {Yang}}, \bibinfo {author} {\bibfnamefont {Z.}~\bibnamefont {Pei}}, \bibinfo
  {author} {\bibfnamefont {E.~C.}\ \bibnamefont {Leon}}, \bibinfo {author}
  {\bibfnamefont {C.}~\bibnamefont {Wickizer}}, \bibinfo {author}
  {\bibfnamefont {B.}~\bibnamefont {Weng}}, \bibinfo {author} {\bibfnamefont
  {Y.}~\bibnamefont {Mao}}, \bibinfo {author} {\bibfnamefont {Q.}~\bibnamefont
  {Ou}},\ and\ \bibinfo {author} {\bibfnamefont {Y.}~\bibnamefont {Shao}},\
  }\bibfield  {title} {\bibinfo {title} {{Cavity quantum-electrodynamical
  time-dependent density functional theory within Gaussian atomic basis. II.
  Analytic energy gradient}},\ }\href {https://doi.org/10.1063/5.0082386}
  {\bibfield  {journal} {\bibinfo  {journal} {The Journal of Chemical Physics}\
  }\textbf {\bibinfo {volume} {156}},\ \bibinfo {pages} {124104} (\bibinfo
  {year} {2022})}\BibitemShut {NoStop}%
\bibitem [{\citenamefont {McTague}\ and\ \citenamefont
  {Foley}(2022)}]{10.1063/5.0091953}%
  \BibitemOpen
  \bibfield  {author} {\bibinfo {author} {\bibfnamefont {J.}~\bibnamefont
  {McTague}}\ and\ \bibinfo {author} {\bibfnamefont {I.}~\bibnamefont {Foley},
  \bibfnamefont {Jonathan~J.}},\ }\bibfield  {title} {\bibinfo {title}
  {{Non-Hermitian cavity quantum electrodynamics–configuration interaction
  singles approach for polaritonic structure with ab initio molecular
  Hamiltonians}},\ }\href {https://doi.org/10.1063/5.0091953} {\bibfield
  {journal} {\bibinfo  {journal} {The Journal of Chemical Physics}\ }\textbf
  {\bibinfo {volume} {156}},\ \bibinfo {pages} {154103} (\bibinfo {year}
  {2022})}\BibitemShut {NoStop}%
\bibitem [{\citenamefont {Bauer}\ and\ \citenamefont
  {Dreuw}(2023)}]{10.1063/5.0142403}%
  \BibitemOpen
  \bibfield  {author} {\bibinfo {author} {\bibfnamefont {M.}~\bibnamefont
  {Bauer}}\ and\ \bibinfo {author} {\bibfnamefont {A.}~\bibnamefont {Dreuw}},\
  }\bibfield  {title} {\bibinfo {title} {{Perturbation theoretical approaches
  to strong light–matter coupling in ground and excited electronic states for
  the description of molecular polaritons}},\ }\href
  {https://doi.org/10.1063/5.0142403} {\bibfield  {journal} {\bibinfo
  {journal} {The Journal of Chemical Physics}\ }\textbf {\bibinfo {volume}
  {158}},\ \bibinfo {pages} {124128} (\bibinfo {year} {2023})}\BibitemShut
  {NoStop}%
\bibitem [{\citenamefont {Cui}\ \emph {et~al.}(2024)\citenamefont {Cui},
  \citenamefont {Mandal},\ and\ \citenamefont
  {Reichman}}]{doi:10.1021/acs.jctc.3c01166}%
  \BibitemOpen
  \bibfield  {author} {\bibinfo {author} {\bibfnamefont {Z.-H.}\ \bibnamefont
  {Cui}}, \bibinfo {author} {\bibfnamefont {A.}~\bibnamefont {Mandal}},\ and\
  \bibinfo {author} {\bibfnamefont {D.~R.}\ \bibnamefont {Reichman}},\
  }\bibfield  {title} {\bibinfo {title} {Variational lang–firsov approach
  plus møller–plesset perturbation theory with applications to ab initio
  polariton chemistry},\ }\href {https://doi.org/10.1021/acs.jctc.3c01166}
  {\bibfield  {journal} {\bibinfo  {journal} {Journal of Chemical Theory and
  Computation}\ }\textbf {\bibinfo {volume} {20}},\ \bibinfo {pages} {1143}
  (\bibinfo {year} {2024})}\BibitemShut {NoStop}%
\bibitem [{\citenamefont {Mallory}\ and\ \citenamefont
  {DePrince}(2022)}]{PhysRevA.106.053710}%
  \BibitemOpen
  \bibfield  {author} {\bibinfo {author} {\bibfnamefont {J.~D.}\ \bibnamefont
  {Mallory}}\ and\ \bibinfo {author} {\bibfnamefont {A.~E.}\ \bibnamefont
  {DePrince}},\ }\bibfield  {title} {\bibinfo {title}
  {Reduced-density-matrix-based ab initio cavity quantum electrodynamics},\
  }\href {https://doi.org/10.1103/PhysRevA.106.053710} {\bibfield  {journal}
  {\bibinfo  {journal} {Phys. Rev. A}\ }\textbf {\bibinfo {volume} {106}},\
  \bibinfo {pages} {053710} (\bibinfo {year} {2022})}\BibitemShut {NoStop}%
\bibitem [{\citenamefont {Weight}\ \emph {et~al.}(2024)\citenamefont {Weight},
  \citenamefont {Tretiak},\ and\ \citenamefont {Zhang}}]{PhysRevA.109.032804}%
  \BibitemOpen
  \bibfield  {author} {\bibinfo {author} {\bibfnamefont {B.~M.}\ \bibnamefont
  {Weight}}, \bibinfo {author} {\bibfnamefont {S.}~\bibnamefont {Tretiak}},\
  and\ \bibinfo {author} {\bibfnamefont {Y.}~\bibnamefont {Zhang}},\ }\bibfield
   {title} {\bibinfo {title} {Diffusion quantum monte carlo approach to the
  polaritonic ground state},\ }\href
  {https://doi.org/10.1103/PhysRevA.109.032804} {\bibfield  {journal} {\bibinfo
   {journal} {Phys. Rev. A}\ }\textbf {\bibinfo {volume} {109}},\ \bibinfo
  {pages} {032804} (\bibinfo {year} {2024})}\BibitemShut {NoStop}%
\bibitem [{\citenamefont {Vu}\ \emph {et~al.}(2024)\citenamefont {Vu},
  \citenamefont {Mejia-Rodriguez}, \citenamefont {Bauman}, \citenamefont
  {Panyala}, \citenamefont {Mutlu}, \citenamefont {Govind},\ and\ \citenamefont
  {Foley~IV}}]{vu2024cavity}%
  \BibitemOpen
  \bibfield  {author} {\bibinfo {author} {\bibfnamefont {N.}~\bibnamefont
  {Vu}}, \bibinfo {author} {\bibfnamefont {D.}~\bibnamefont {Mejia-Rodriguez}},
  \bibinfo {author} {\bibfnamefont {N.~P.}\ \bibnamefont {Bauman}}, \bibinfo
  {author} {\bibfnamefont {A.}~\bibnamefont {Panyala}}, \bibinfo {author}
  {\bibfnamefont {E.}~\bibnamefont {Mutlu}}, \bibinfo {author} {\bibfnamefont
  {N.}~\bibnamefont {Govind}},\ and\ \bibinfo {author} {\bibfnamefont {J.~J.}\
  \bibnamefont {Foley~IV}},\ }\bibfield  {title} {\bibinfo {title} {Cavity
  quantum electrodynamics complete active space configuration interaction
  theory},\ }\href@noop {} {\bibfield  {journal} {\bibinfo  {journal} {Journal
  of Chemical Theory and Computation}\ } (\bibinfo {year} {2024})}\BibitemShut
  {NoStop}%
\bibitem [{\citenamefont {Coester}(1958)}]{coester58_421}%
  \BibitemOpen
  \bibfield  {author} {\bibinfo {author} {\bibfnamefont {F.}~\bibnamefont
  {Coester}},\ }\bibfield  {title} {\bibinfo {title} {Bound states of a
  many-particle system},\ }\href
  {https://doi.org/http://dx.doi.org/10.1016/0029-5582(58)90280-3} {\bibfield
  {journal} {\bibinfo  {journal} {Nucl. Phys.}\ }\textbf {\bibinfo {volume}
  {7}},\ \bibinfo {pages} {421} (\bibinfo {year} {1958})}\BibitemShut {NoStop}%
\bibitem [{\citenamefont {Coester}\ and\ \citenamefont
  {K{\"u}mmel}(1960)}]{coester60_477}%
  \BibitemOpen
  \bibfield  {author} {\bibinfo {author} {\bibfnamefont {F.}~\bibnamefont
  {Coester}}\ and\ \bibinfo {author} {\bibfnamefont {H.}~\bibnamefont
  {K{\"u}mmel}},\ }\bibfield  {title} {\bibinfo {title} {Short-range
  correlations in nuclear wave functions},\ }\href
  {https://doi.org/http://dx.doi.org/10.1016/0029-5582(60)90140-1} {\bibfield
  {journal} {\bibinfo  {journal} {Nucl. Phys.}\ }\textbf {\bibinfo {volume}
  {17}},\ \bibinfo {pages} {477} (\bibinfo {year} {1960})}\BibitemShut
  {NoStop}%
\bibitem [{\citenamefont {{\v C}{\'\i}{\v z}ek}(1966)}]{cizek66_4256}%
  \BibitemOpen
  \bibfield  {author} {\bibinfo {author} {\bibfnamefont {J.}~\bibnamefont {{\v
  C}{\'\i}{\v z}ek}},\ }\bibfield  {title} {\bibinfo {title} {On the
  correlation problem in atomic and molecular systems. calculation of
  wavefunction components in ursell-type expansion using quantum-field
  theoretical methods},\ }\href
  {https://doi.org/http://dx.doi.org/10.1063/1.1727484} {\bibfield  {journal}
  {\bibinfo  {journal} {J. Chem. Phys.}\ }\textbf {\bibinfo {volume} {45}},\
  \bibinfo {pages} {4256} (\bibinfo {year} {1966})}\BibitemShut {NoStop}%
\bibitem [{\citenamefont {Paldus}\ \emph {et~al.}(1972)\citenamefont {Paldus},
  \citenamefont {\ifmmode \check{C}\else
  \v{C}\fi{}\'{\i}\ifmmode~\check{z}\else \v{z}\fi{}ek},\ and\ \citenamefont
  {Shavitt}}]{paldus72_50}%
  \BibitemOpen
  \bibfield  {author} {\bibinfo {author} {\bibfnamefont {J.}~\bibnamefont
  {Paldus}}, \bibinfo {author} {\bibfnamefont {J.}~\bibnamefont {\ifmmode
  \check{C}\else \v{C}\fi{}\'{\i}\ifmmode~\check{z}\else \v{z}\fi{}ek}},\ and\
  \bibinfo {author} {\bibfnamefont {I.}~\bibnamefont {Shavitt}},\ }\bibfield
  {title} {\bibinfo {title} {Correlation problems in atomic and molecular
  systems. iv. extended coupled-pair many-electron theory and its application
  to the b${\mathrm{h}}_{3}$ molecule},\ }\href
  {https://doi.org/10.1103/PhysRevA.5.50} {\bibfield  {journal} {\bibinfo
  {journal} {Phys. Rev. A}\ }\textbf {\bibinfo {volume} {5}},\ \bibinfo {pages}
  {50} (\bibinfo {year} {1972})}\BibitemShut {NoStop}%
\bibitem [{\citenamefont {K{\"u}mmel}(2003)}]{kummel2003biography}%
  \BibitemOpen
  \bibfield  {author} {\bibinfo {author} {\bibfnamefont {H.~G.}\ \bibnamefont
  {K{\"u}mmel}},\ }\bibfield  {title} {\bibinfo {title} {A biography of the
  coupled cluster method},\ }\href@noop {} {\bibfield  {journal} {\bibinfo
  {journal} {Int. J. Mod. Phys. B}\ }\textbf {\bibinfo {volume} {17}},\
  \bibinfo {pages} {5311} (\bibinfo {year} {2003})}\BibitemShut {NoStop}%
\bibitem [{\citenamefont {Bartlett}\ and\ \citenamefont
  {Musia\l{}}(2007)}]{RevModPhys.79.291}%
  \BibitemOpen
  \bibfield  {author} {\bibinfo {author} {\bibfnamefont {R.~J.}\ \bibnamefont
  {Bartlett}}\ and\ \bibinfo {author} {\bibfnamefont {M.}~\bibnamefont
  {Musia\l{}}},\ }\bibfield  {title} {\bibinfo {title} {Coupled-cluster theory
  in quantum chemistry},\ }\href {https://doi.org/10.1103/RevModPhys.79.291}
  {\bibfield  {journal} {\bibinfo  {journal} {Rev. Mod. Phys.}\ }\textbf
  {\bibinfo {volume} {79}},\ \bibinfo {pages} {291} (\bibinfo {year}
  {2007})}\BibitemShut {NoStop}%
\bibitem [{\citenamefont {Mukherjee}\ and\ \citenamefont
  {Pal}(1989)}]{mukherjee1989use}%
  \BibitemOpen
  \bibfield  {author} {\bibinfo {author} {\bibfnamefont {D.}~\bibnamefont
  {Mukherjee}}\ and\ \bibinfo {author} {\bibfnamefont {S.}~\bibnamefont
  {Pal}},\ }\bibfield  {title} {\bibinfo {title} {Use of cluster expansion
  methods in the open-shell correlation problem},\ }\href@noop {} {\bibfield
  {journal} {\bibinfo  {journal} {Adv. Quantum Chem.}\ }\textbf {\bibinfo
  {volume} {20}},\ \bibinfo {pages} {291} (\bibinfo {year} {1989})}\BibitemShut
  {NoStop}%
\bibitem [{\citenamefont {Krylov}(2008)}]{krylov2008equation}%
  \BibitemOpen
  \bibfield  {author} {\bibinfo {author} {\bibfnamefont {A.~I.}\ \bibnamefont
  {Krylov}},\ }\bibfield  {title} {\bibinfo {title} {Equation-of-motion
  coupled-cluster methods for open-shell and electronically excited species:
  the hitchhiker's guide to fock space},\ }\href@noop {} {\bibfield  {journal}
  {\bibinfo  {journal} {Annu. Rev. Phys. Chem.}\ }\textbf {\bibinfo {volume}
  {59}},\ \bibinfo {pages} {433} (\bibinfo {year} {2008})}\BibitemShut
  {NoStop}%
\bibitem [{\citenamefont {Piecuch}\ \emph {et~al.}(2002)\citenamefont
  {Piecuch}, \citenamefont {Kowalski}, \citenamefont {Pimienta},\ and\
  \citenamefont {Mcguire}}]{piecuch2002recent}%
  \BibitemOpen
  \bibfield  {author} {\bibinfo {author} {\bibfnamefont {P.}~\bibnamefont
  {Piecuch}}, \bibinfo {author} {\bibfnamefont {K.}~\bibnamefont {Kowalski}},
  \bibinfo {author} {\bibfnamefont {I.~S.}\ \bibnamefont {Pimienta}},\ and\
  \bibinfo {author} {\bibfnamefont {M.~J.}\ \bibnamefont {Mcguire}},\
  }\bibfield  {title} {\bibinfo {title} {Recent advances in electronic
  structure theory: Method of moments of coupled-cluster equations and
  renormalized coupled-cluster approaches},\ }\href@noop {} {\bibfield
  {journal} {\bibinfo  {journal} {Int. Rev. Phys. Chem.}\ }\textbf {\bibinfo
  {volume} {21}},\ \bibinfo {pages} {527} (\bibinfo {year} {2002})}\BibitemShut
  {NoStop}%
\bibitem [{\citenamefont {Crawford}\ and\ \citenamefont
  {Schaefer}(2000)}]{crawford2000introduction}%
  \BibitemOpen
  \bibfield  {author} {\bibinfo {author} {\bibfnamefont {T.~D.}\ \bibnamefont
  {Crawford}}\ and\ \bibinfo {author} {\bibfnamefont {H.~F.}\ \bibnamefont
  {Schaefer}},\ }\bibfield  {title} {\bibinfo {title} {An introduction to
  coupled cluster theory for computational chemists},\ }\href@noop {}
  {\bibfield  {journal} {\bibinfo  {journal} {Rev. Comput. Chem.}\ }\textbf
  {\bibinfo {volume} {14}},\ \bibinfo {pages} {33} (\bibinfo {year}
  {2000})}\BibitemShut {NoStop}%
\bibitem [{\citenamefont {Pavo\v{s}evi\'c}\ \emph {et~al.}(2022)\citenamefont
  {Pavo\v{s}evi\'c}, \citenamefont {Hammes-Schiffer}, \citenamefont {Rubio},\
  and\ \citenamefont {Flick}}]{flick_jacs}%
  \BibitemOpen
  \bibfield  {author} {\bibinfo {author} {\bibfnamefont {F.}~\bibnamefont
  {Pavo\v{s}evi\'c}}, \bibinfo {author} {\bibfnamefont {S.}~\bibnamefont
  {Hammes-Schiffer}}, \bibinfo {author} {\bibfnamefont {A.}~\bibnamefont
  {Rubio}},\ and\ \bibinfo {author} {\bibfnamefont {J.}~\bibnamefont {Flick}},\
  }\bibfield  {title} {\bibinfo {title} {Cavity-modulated proton transfer
  reactions},\ }\href {https://doi.org/10.1021/jacs.1c13201} {\bibfield
  {journal} {\bibinfo  {journal} {Journal of the American Chemical Society}\
  }\textbf {\bibinfo {volume} {144}},\ \bibinfo {pages} {4995} (\bibinfo {year}
  {2022})}\BibitemShut {NoStop}%
\bibitem [{\citenamefont {Mordovina}\ \emph {et~al.}(2020)\citenamefont
  {Mordovina}, \citenamefont {Bungey}, \citenamefont {Appel}, \citenamefont
  {Knowles}, \citenamefont {Rubio},\ and\ \citenamefont {Manby}}]{mordovina}%
  \BibitemOpen
  \bibfield  {author} {\bibinfo {author} {\bibfnamefont {U.}~\bibnamefont
  {Mordovina}}, \bibinfo {author} {\bibfnamefont {C.}~\bibnamefont {Bungey}},
  \bibinfo {author} {\bibfnamefont {H.}~\bibnamefont {Appel}}, \bibinfo
  {author} {\bibfnamefont {P.~J.}\ \bibnamefont {Knowles}}, \bibinfo {author}
  {\bibfnamefont {A.}~\bibnamefont {Rubio}},\ and\ \bibinfo {author}
  {\bibfnamefont {F.~R.}\ \bibnamefont {Manby}},\ }\bibfield  {title} {\bibinfo
  {title} {Polaritonic coupled-cluster theory},\ }\href
  {https://doi.org/10.1103/PhysRevResearch.2.023262} {\bibfield  {journal}
  {\bibinfo  {journal} {Phys. Rev. Res.}\ }\textbf {\bibinfo {volume} {2}},\
  \bibinfo {pages} {023262} (\bibinfo {year} {2020})}\BibitemShut {NoStop}%
\bibitem [{\citenamefont {Haugland}\ \emph {et~al.}(2020)\citenamefont
  {Haugland}, \citenamefont {Ronca}, \citenamefont {Kj\o{}nstad}, \citenamefont
  {Rubio},\ and\ \citenamefont {Koch}}]{rubio_prx}%
  \BibitemOpen
  \bibfield  {author} {\bibinfo {author} {\bibfnamefont {T.~S.}\ \bibnamefont
  {Haugland}}, \bibinfo {author} {\bibfnamefont {E.}~\bibnamefont {Ronca}},
  \bibinfo {author} {\bibfnamefont {E.~F.}\ \bibnamefont {Kj\o{}nstad}},
  \bibinfo {author} {\bibfnamefont {A.}~\bibnamefont {Rubio}},\ and\ \bibinfo
  {author} {\bibfnamefont {H.}~\bibnamefont {Koch}},\ }\bibfield  {title}
  {\bibinfo {title} {Coupled cluster theory for molecular polaritons: Changing
  ground and excited states},\ }\href
  {https://doi.org/10.1103/PhysRevX.10.041043} {\bibfield  {journal} {\bibinfo
  {journal} {Phys. Rev. X}\ }\textbf {\bibinfo {volume} {10}},\ \bibinfo
  {pages} {041043} (\bibinfo {year} {2020})}\BibitemShut {NoStop}%
\bibitem [{\citenamefont {DePrince}(2021)}]{10.1063/5.0038748}%
  \BibitemOpen
  \bibfield  {author} {\bibinfo {author} {\bibfnamefont {I.}~\bibnamefont
  {DePrince}, \bibfnamefont {A.~Eugene}},\ }\bibfield  {title} {\bibinfo
  {title} {{Cavity-modulated ionization potentials and electron affinities from
  quantum electrodynamics coupled-cluster theory}},\ }\href
  {https://doi.org/10.1063/5.0038748} {\bibfield  {journal} {\bibinfo
  {journal} {The Journal of Chemical Physics}\ }\textbf {\bibinfo {volume}
  {154}},\ \bibinfo {pages} {094112} (\bibinfo {year} {2021})}\BibitemShut
  {NoStop}%
\bibitem [{\citenamefont {Pavo\v{s}evi\'c}\ and\ \citenamefont
  {Flick}(2021)}]{doi:10.1021/acs.jpclett.1c02659}%
  \BibitemOpen
  \bibfield  {author} {\bibinfo {author} {\bibfnamefont {F.}~\bibnamefont
  {Pavo\v{s}evi\'c}}\ and\ \bibinfo {author} {\bibfnamefont {J.}~\bibnamefont
  {Flick}},\ }\bibfield  {title} {\bibinfo {title} {Polaritonic unitary coupled
  cluster for quantum computations},\ }\href
  {https://doi.org/10.1021/acs.jpclett.1c02659} {\bibfield  {journal} {\bibinfo
   {journal} {The Journal of Physical Chemistry Letters}\ }\textbf {\bibinfo
  {volume} {12}},\ \bibinfo {pages} {9100} (\bibinfo {year}
  {2021})}\BibitemShut {NoStop}%
\bibitem [{\citenamefont {Tilly}\ \emph {et~al.}(2022)\citenamefont {Tilly},
  \citenamefont {Chen}, \citenamefont {Cao}, \citenamefont {Picozzi},
  \citenamefont {Setia}, \citenamefont {Li}, \citenamefont {Grant},
  \citenamefont {Wossnig}, \citenamefont {Rungger}, \citenamefont {Booth} \emph
  {et~al.}}]{tilly2022variational}%
  \BibitemOpen
  \bibfield  {author} {\bibinfo {author} {\bibfnamefont {J.}~\bibnamefont
  {Tilly}}, \bibinfo {author} {\bibfnamefont {H.}~\bibnamefont {Chen}},
  \bibinfo {author} {\bibfnamefont {S.}~\bibnamefont {Cao}}, \bibinfo {author}
  {\bibfnamefont {D.}~\bibnamefont {Picozzi}}, \bibinfo {author} {\bibfnamefont
  {K.}~\bibnamefont {Setia}}, \bibinfo {author} {\bibfnamefont
  {Y.}~\bibnamefont {Li}}, \bibinfo {author} {\bibfnamefont {E.}~\bibnamefont
  {Grant}}, \bibinfo {author} {\bibfnamefont {L.}~\bibnamefont {Wossnig}},
  \bibinfo {author} {\bibfnamefont {I.}~\bibnamefont {Rungger}}, \bibinfo
  {author} {\bibfnamefont {G.~H.}\ \bibnamefont {Booth}}, \emph {et~al.},\
  }\bibfield  {title} {\bibinfo {title} {The variational quantum eigensolver: a
  review of methods and best practices},\ }\href
  {https://doi.org/https://doi.org/10.1016/j.physrep.2022.08.003} {\bibfield
  {journal} {\bibinfo  {journal} {Physics Reports}\ }\textbf {\bibinfo {volume}
  {986}},\ \bibinfo {pages} {1} (\bibinfo {year} {2022})}\BibitemShut {NoStop}%
\bibitem [{\citenamefont {Riso}\ \emph {et~al.}(2022)\citenamefont {Riso},
  \citenamefont {Haugland}, \citenamefont {Ronca},\ and\ \citenamefont
  {Koch}}]{10.1063/5.0091119}%
  \BibitemOpen
  \bibfield  {author} {\bibinfo {author} {\bibfnamefont {R.~R.}\ \bibnamefont
  {Riso}}, \bibinfo {author} {\bibfnamefont {T.~S.}\ \bibnamefont {Haugland}},
  \bibinfo {author} {\bibfnamefont {E.}~\bibnamefont {Ronca}},\ and\ \bibinfo
  {author} {\bibfnamefont {H.}~\bibnamefont {Koch}},\ }\bibfield  {title}
  {\bibinfo {title} {{On the characteristic features of ionization in QED
  environments}},\ }\href {https://doi.org/10.1063/5.0091119} {\bibfield
  {journal} {\bibinfo  {journal} {The Journal of Chemical Physics}\ }\textbf
  {\bibinfo {volume} {156}},\ \bibinfo {pages} {234103} (\bibinfo {year}
  {2022})}\BibitemShut {NoStop}%
\bibitem [{\citenamefont {White}\ \emph {et~al.}(2020)\citenamefont {White},
  \citenamefont {Gao}, \citenamefont {Minnich},\ and\ \citenamefont
  {Chan}}]{white}%
  \BibitemOpen
  \bibfield  {author} {\bibinfo {author} {\bibfnamefont {A.~F.}\ \bibnamefont
  {White}}, \bibinfo {author} {\bibfnamefont {Y.}~\bibnamefont {Gao}}, \bibinfo
  {author} {\bibfnamefont {A.~J.}\ \bibnamefont {Minnich}},\ and\ \bibinfo
  {author} {\bibfnamefont {G.~K.-L.}\ \bibnamefont {Chan}},\ }\bibfield
  {title} {\bibinfo {title} {{A coupled cluster framework for electrons and
  phonons}},\ }\href {https://doi.org/10.1063/5.0033132} {\bibfield  {journal}
  {\bibinfo  {journal} {The Journal of Chemical Physics}\ }\textbf {\bibinfo
  {volume} {153}},\ \bibinfo {pages} {224112} (\bibinfo {year}
  {2020})}\BibitemShut {NoStop}%
\bibitem [{\citenamefont {Foley}\ \emph {et~al.}(2023)\citenamefont {Foley},
  \citenamefont {McTague},\ and\ \citenamefont {DePrince}}]{deprince_review}%
  \BibitemOpen
  \bibfield  {author} {\bibinfo {author} {\bibfnamefont {I.}~\bibnamefont
  {Foley}, \bibfnamefont {Jonathan~J.}}, \bibinfo {author} {\bibfnamefont
  {J.~F.}\ \bibnamefont {McTague}},\ and\ \bibinfo {author} {\bibfnamefont
  {I.}~\bibnamefont {DePrince}, \bibfnamefont {A.~Eugene}},\ }\bibfield
  {title} {\bibinfo {title} {{Ab initio methods for polariton chemistry}},\
  }\href {https://doi.org/10.1063/5.0167243} {\bibfield  {journal} {\bibinfo
  {journal} {Chemical Physics Reviews}\ }\textbf {\bibinfo {volume} {4}},\
  \bibinfo {pages} {041301} (\bibinfo {year} {2023})}\BibitemShut {NoStop}%
\bibitem [{\citenamefont {Panyala}\ \emph {et~al.}(2023)\citenamefont
  {Panyala}, \citenamefont {Govind}, \citenamefont {Kowalski}, \citenamefont
  {Bauman}, \citenamefont {Peng}, \citenamefont {Pathak}, \citenamefont
  {Mutlu}, \citenamefont {Mejia~Rodriguez}, \citenamefont {Xantheas},\ and\
  \citenamefont {Krishnamoorthy}}]{doecode_108784}%
  \BibitemOpen
  \bibfield  {author} {\bibinfo {author} {\bibfnamefont {A.}~\bibnamefont
  {Panyala}}, \bibinfo {author} {\bibfnamefont {N.}~\bibnamefont {Govind}},
  \bibinfo {author} {\bibfnamefont {K.}~\bibnamefont {Kowalski}}, \bibinfo
  {author} {\bibfnamefont {N.}~\bibnamefont {Bauman}}, \bibinfo {author}
  {\bibfnamefont {B.}~\bibnamefont {Peng}}, \bibinfo {author} {\bibfnamefont
  {H.}~\bibnamefont {Pathak}}, \bibinfo {author} {\bibfnamefont
  {E.}~\bibnamefont {Mutlu}}, \bibinfo {author} {\bibfnamefont
  {D.}~\bibnamefont {Mejia~Rodriguez}}, \bibinfo {author} {\bibfnamefont
  {S.}~\bibnamefont {Xantheas}},\ and\ \bibinfo {author} {\bibfnamefont
  {S.}~\bibnamefont {Krishnamoorthy}},\ }\href
  {https://doi.org/10.11578/dc.20230628.1} {\bibinfo {title}
  {Exachem/exachem}},\ \bibinfo {howpublished} {[Computer Software]
  \url{https://doi.org/10.11578/dc.20230628.1}} (\bibinfo {year}
  {2023})\BibitemShut {NoStop}%
\bibitem [{\citenamefont {Mutlu}\ \emph {et~al.}(2023)\citenamefont {Mutlu},
  \citenamefont {Panyala}, \citenamefont {Gawande}, \citenamefont {Bagusetty},
  \citenamefont {Glabe}, \citenamefont {Kim}, \citenamefont {Kowalski},
  \citenamefont {Bauman}, \citenamefont {Peng}, \citenamefont {Pathak},
  \citenamefont {Brabec},\ and\ \citenamefont {Krishnamoorthy}}]{tamm}%
  \BibitemOpen
  \bibfield  {author} {\bibinfo {author} {\bibfnamefont {E.}~\bibnamefont
  {Mutlu}}, \bibinfo {author} {\bibfnamefont {A.}~\bibnamefont {Panyala}},
  \bibinfo {author} {\bibfnamefont {N.}~\bibnamefont {Gawande}}, \bibinfo
  {author} {\bibfnamefont {A.}~\bibnamefont {Bagusetty}}, \bibinfo {author}
  {\bibfnamefont {J.}~\bibnamefont {Glabe}}, \bibinfo {author} {\bibfnamefont
  {J.}~\bibnamefont {Kim}}, \bibinfo {author} {\bibfnamefont {K.}~\bibnamefont
  {Kowalski}}, \bibinfo {author} {\bibfnamefont {N.~P.}\ \bibnamefont
  {Bauman}}, \bibinfo {author} {\bibfnamefont {B.}~\bibnamefont {Peng}},
  \bibinfo {author} {\bibfnamefont {H.}~\bibnamefont {Pathak}}, \bibinfo
  {author} {\bibfnamefont {J.}~\bibnamefont {Brabec}},\ and\ \bibinfo {author}
  {\bibfnamefont {S.}~\bibnamefont {Krishnamoorthy}},\ }\bibfield  {title}
  {\bibinfo {title} {{TAMM: Tensor algebra for many-body methods}},\ }\href
  {https://doi.org/10.1063/5.0142433} {\bibfield  {journal} {\bibinfo
  {journal} {The Journal of Chemical Physics}\ }\textbf {\bibinfo {volume}
  {159}},\ \bibinfo {pages} {024801} (\bibinfo {year} {2023})}\BibitemShut
  {NoStop}%
\bibitem [{\citenamefont {Cohen-Tannoudji}\ \emph {et~al.}(1997)\citenamefont
  {Cohen-Tannoudji}, \citenamefont {Dupont-Roc},\ and\ \citenamefont
  {Grynberg}}]{cohen1997photons}%
  \BibitemOpen
  \bibfield  {author} {\bibinfo {author} {\bibfnamefont {C.}~\bibnamefont
  {Cohen-Tannoudji}}, \bibinfo {author} {\bibfnamefont {J.}~\bibnamefont
  {Dupont-Roc}},\ and\ \bibinfo {author} {\bibfnamefont {G.}~\bibnamefont
  {Grynberg}},\ }\href@noop {} {\emph {\bibinfo {title} {Photons and
  atoms-introduction to quantum electrodynamics}}}\ (\bibinfo {year}
  {1997})\BibitemShut {NoStop}%
\bibitem [{\citenamefont {Spohn}(2004)}]{spohn2004dynamics}%
  \BibitemOpen
  \bibfield  {author} {\bibinfo {author} {\bibfnamefont {H.}~\bibnamefont
  {Spohn}},\ }\href@noop {} {\emph {\bibinfo {title} {Dynamics of charged
  particles and their radiation field}}}\ (\bibinfo  {publisher} {Cambridge
  university press},\ \bibinfo {year} {2004})\BibitemShut {NoStop}%
\bibitem [{\citenamefont {Rokaj}\ \emph {et~al.}(2018)\citenamefont {Rokaj},
  \citenamefont {Welakuh}, \citenamefont {Ruggenthaler},\ and\ \citenamefont
  {Rubio}}]{rokaj2018light}%
  \BibitemOpen
  \bibfield  {author} {\bibinfo {author} {\bibfnamefont {V.}~\bibnamefont
  {Rokaj}}, \bibinfo {author} {\bibfnamefont {D.~M.}\ \bibnamefont {Welakuh}},
  \bibinfo {author} {\bibfnamefont {M.}~\bibnamefont {Ruggenthaler}},\ and\
  \bibinfo {author} {\bibfnamefont {A.}~\bibnamefont {Rubio}},\ }\bibfield
  {title} {\bibinfo {title} {Light--matter interaction in the long-wavelength
  limit: no ground-state without dipole self-energy},\ }\href@noop {}
  {\bibfield  {journal} {\bibinfo  {journal} {Journal of Physics B: Atomic,
  Molecular and Optical Physics}\ }\textbf {\bibinfo {volume} {51}},\ \bibinfo
  {pages} {034005} (\bibinfo {year} {2018})}\BibitemShut {NoStop}%
\bibitem [{\citenamefont {Climent}\ \emph {et~al.}(2019)\citenamefont
  {Climent}, \citenamefont {Galego}, \citenamefont {Garcia-Vidal},\ and\
  \citenamefont {Feist}}]{climent2019plasmonic}%
  \BibitemOpen
  \bibfield  {author} {\bibinfo {author} {\bibfnamefont {C.}~\bibnamefont
  {Climent}}, \bibinfo {author} {\bibfnamefont {J.}~\bibnamefont {Galego}},
  \bibinfo {author} {\bibfnamefont {F.~J.}\ \bibnamefont {Garcia-Vidal}},\ and\
  \bibinfo {author} {\bibfnamefont {J.}~\bibnamefont {Feist}},\ }\bibfield
  {title} {\bibinfo {title} {Plasmonic nanocavities enable self-induced
  electrostatic catalysis},\ }\href@noop {} {\bibfield  {journal} {\bibinfo
  {journal} {Angewandte Chemie International Edition}\ }\textbf {\bibinfo
  {volume} {58}},\ \bibinfo {pages} {8698} (\bibinfo {year}
  {2019})}\BibitemShut {NoStop}%
\bibitem [{\citenamefont {Sch\"{a}fer}\ \emph {et~al.}(2020)\citenamefont
  {Sch\"{a}fer}, \citenamefont {Ruggenthaler}, \citenamefont {Rokaj},\ and\
  \citenamefont {Rubio}}]{doi:10.1021/acsphotonics.9b01649}%
  \BibitemOpen
  \bibfield  {author} {\bibinfo {author} {\bibfnamefont {C.}~\bibnamefont
  {Sch\"{a}fer}}, \bibinfo {author} {\bibfnamefont {M.}~\bibnamefont
  {Ruggenthaler}}, \bibinfo {author} {\bibfnamefont {V.}~\bibnamefont
  {Rokaj}},\ and\ \bibinfo {author} {\bibfnamefont {A.}~\bibnamefont {Rubio}},\
  }\bibfield  {title} {\bibinfo {title} {Relevance of the quadratic diamagnetic
  and self-polarization terms in cavity quantum electrodynamics},\ }\href
  {https://doi.org/10.1021/acsphotonics.9b01649} {\bibfield  {journal}
  {\bibinfo  {journal} {ACS Photonics}\ }\textbf {\bibinfo {volume} {7}},\
  \bibinfo {pages} {975} (\bibinfo {year} {2020})}\BibitemShut {NoStop}%
\bibitem [{\citenamefont {De~Bernardis}\ \emph {et~al.}(2018)\citenamefont
  {De~Bernardis}, \citenamefont {Jaako},\ and\ \citenamefont
  {Rabl}}]{PhysRevA.97.043820}%
  \BibitemOpen
  \bibfield  {author} {\bibinfo {author} {\bibfnamefont {D.}~\bibnamefont
  {De~Bernardis}}, \bibinfo {author} {\bibfnamefont {T.}~\bibnamefont
  {Jaako}},\ and\ \bibinfo {author} {\bibfnamefont {P.}~\bibnamefont {Rabl}},\
  }\bibfield  {title} {\bibinfo {title} {Cavity quantum electrodynamics in the
  nonperturbative regime},\ }\href {https://doi.org/10.1103/PhysRevA.97.043820}
  {\bibfield  {journal} {\bibinfo  {journal} {Phys. Rev. A}\ }\textbf {\bibinfo
  {volume} {97}},\ \bibinfo {pages} {043820} (\bibinfo {year}
  {2018})}\BibitemShut {NoStop}%
\bibitem [{\citenamefont {Taylor}\ \emph {et~al.}(2020)\citenamefont {Taylor},
  \citenamefont {Mandal}, \citenamefont {Zhou},\ and\ \citenamefont
  {Huo}}]{PhysRevLett.125.123602}%
  \BibitemOpen
  \bibfield  {author} {\bibinfo {author} {\bibfnamefont {M.~A.~D.}\
  \bibnamefont {Taylor}}, \bibinfo {author} {\bibfnamefont {A.}~\bibnamefont
  {Mandal}}, \bibinfo {author} {\bibfnamefont {W.}~\bibnamefont {Zhou}},\ and\
  \bibinfo {author} {\bibfnamefont {P.}~\bibnamefont {Huo}},\ }\bibfield
  {title} {\bibinfo {title} {Resolution of gauge ambiguities in molecular
  cavity quantum electrodynamics},\ }\href
  {https://doi.org/10.1103/PhysRevLett.125.123602} {\bibfield  {journal}
  {\bibinfo  {journal} {Phys. Rev. Lett.}\ }\textbf {\bibinfo {volume} {125}},\
  \bibinfo {pages} {123602} (\bibinfo {year} {2020})}\BibitemShut {NoStop}%
\bibitem [{\citenamefont {Rivera}\ \emph {et~al.}(2019)\citenamefont {Rivera},
  \citenamefont {Flick},\ and\ \citenamefont {Narang}}]{rivera2019variational}%
  \BibitemOpen
  \bibfield  {author} {\bibinfo {author} {\bibfnamefont {N.}~\bibnamefont
  {Rivera}}, \bibinfo {author} {\bibfnamefont {J.}~\bibnamefont {Flick}},\ and\
  \bibinfo {author} {\bibfnamefont {P.}~\bibnamefont {Narang}},\ }\bibfield
  {title} {\bibinfo {title} {Variational theory of nonrelativistic quantum
  electrodynamics},\ }\href@noop {} {\bibfield  {journal} {\bibinfo  {journal}
  {Physical review letters}\ }\textbf {\bibinfo {volume} {122}},\ \bibinfo
  {pages} {193603} (\bibinfo {year} {2019})}\BibitemShut {NoStop}%
\bibitem [{\citenamefont {Bartlett}\ and\ \citenamefont
  {Musia{\l}}(2007)}]{bartlett2007coupled}%
  \BibitemOpen
  \bibfield  {author} {\bibinfo {author} {\bibfnamefont {R.~J.}\ \bibnamefont
  {Bartlett}}\ and\ \bibinfo {author} {\bibfnamefont {M.}~\bibnamefont
  {Musia{\l}}},\ }\bibfield  {title} {\bibinfo {title} {Coupled-cluster theory
  in quantum chemistry},\ }\href@noop {} {\bibfield  {journal} {\bibinfo
  {journal} {Reviews of Modern Physics}\ }\textbf {\bibinfo {volume} {79}},\
  \bibinfo {pages} {291} (\bibinfo {year} {2007})}\BibitemShut {NoStop}%
\bibitem [{\citenamefont {Nieplocha}\ \emph {et~al.}(2006)\citenamefont
  {Nieplocha}, \citenamefont {Palmer}, \citenamefont {Tipparaju}, \citenamefont
  {Krishnan}, \citenamefont {Trease},\ and\ \citenamefont {Aprà}}]{ga_paper}%
  \BibitemOpen
  \bibfield  {author} {\bibinfo {author} {\bibfnamefont {J.}~\bibnamefont
  {Nieplocha}}, \bibinfo {author} {\bibfnamefont {B.}~\bibnamefont {Palmer}},
  \bibinfo {author} {\bibfnamefont {V.}~\bibnamefont {Tipparaju}}, \bibinfo
  {author} {\bibfnamefont {M.}~\bibnamefont {Krishnan}}, \bibinfo {author}
  {\bibfnamefont {H.}~\bibnamefont {Trease}},\ and\ \bibinfo {author}
  {\bibfnamefont {E.}~\bibnamefont {Aprà}},\ }\bibfield  {title} {\bibinfo
  {title} {Advances, applications and performance of the global arrays shared
  memory programming toolkit},\ }\href
  {https://doi.org/10.1177/1094342006064503} {\bibfield  {journal} {\bibinfo
  {journal} {The International Journal of High Performance Computing
  Applications}\ }\textbf {\bibinfo {volume} {20}},\ \bibinfo {pages} {203}
  (\bibinfo {year} {2006})}\BibitemShut {NoStop}%
\bibitem [{\citenamefont {Dunning~Jr}(1989)}]{dunning1989gaussian}%
  \BibitemOpen
  \bibfield  {author} {\bibinfo {author} {\bibfnamefont {T.~H.}\ \bibnamefont
  {Dunning~Jr}},\ }\bibfield  {title} {\bibinfo {title} {Gaussian basis sets
  for use in correlated molecular calculations. i. the atoms boron through neon
  and hydrogen},\ }\href@noop {} {\bibfield  {journal} {\bibinfo  {journal}
  {The Journal of chemical physics}\ }\textbf {\bibinfo {volume} {90}},\
  \bibinfo {pages} {1007} (\bibinfo {year} {1989})}\BibitemShut {NoStop}%
\bibitem [{\citenamefont {Kendall}\ \emph {et~al.}(1992)\citenamefont
  {Kendall}, \citenamefont {Dunning},\ and\ \citenamefont
  {Harrison}}]{kendall1992electron}%
  \BibitemOpen
  \bibfield  {author} {\bibinfo {author} {\bibfnamefont {R.~A.}\ \bibnamefont
  {Kendall}}, \bibinfo {author} {\bibfnamefont {T.~H.}\ \bibnamefont
  {Dunning}},\ and\ \bibinfo {author} {\bibfnamefont {R.~J.}\ \bibnamefont
  {Harrison}},\ }\bibfield  {title} {\bibinfo {title} {Electron affinities of
  the first-row atoms revisited. systematic basis sets and wave functions},\
  }\href@noop {} {\bibfield  {journal} {\bibinfo  {journal} {The Journal of
  chemical physics}\ }\textbf {\bibinfo {volume} {96}},\ \bibinfo {pages}
  {6796} (\bibinfo {year} {1992})}\BibitemShut {NoStop}%
\end{thebibliography}
\end{document}